\documentclass[12pt]{article}

\usepackage[margin=1in]{geometry}
\usepackage{graphicx}
\usepackage{caption}
\usepackage{subcaption}
\usepackage{stmaryrd}
\usepackage{xfrac}
\usepackage{braket}

\usepackage{soul}

\usepackage{etoolbox}
\patchcmd{\thebibliography}
  {\settowidth}
  {\setlength{\itemsep}{0pt plus 0.1pt}\settowidth}
  {}{}
\apptocmd{\thebibliography}
  {\small}
  {}{}

\usepackage{amsmath}
\usepackage{amssymb}
\usepackage{tikz}
\usepackage{verbatim}
\usetikzlibrary{arrows,trees,decorations.markings,patterns}

\usepackage{textcomp}

 \newtheorem{@definition}{\bf Definition}[section]

 \newtheorem{@example}{\bf Example}[section]

\newtheorem{@nonexample}{\bf (Non)Example}[section]

 \newtheorem{@remark}{\bf Remark}[section]

\newcommand{\pp}{\mathbb{P}}

\begin{document}

\begin{center}
{\Large \bf Ruling out Bipartite Nonsignaling Nonlocal Models for Tripartite Correlations}
\end{center}

\begin{center}
Peter Bierhorst\footnote{Department of Mathematics, University of New Orleans, Louisiana, USA} 
\end{center}

\begin{abstract}
Many three-party correlations, including some that are commonly described as genuinely tripartite nonlocal, can be simulated by a network of underlying subsystems that display only bipartite nonsignaling nonlocal behavior. Quantum mechanics predicts three-party correlations that admit no such simulation, suggesting there are versions of nonlocality in nature transcending the phenomenon of bipartite nonsignaling nonlocality. This paper introduces a rigorous framework for analyzing tripartite correlations that can be simulated by bipartite-only networks. We confirm that expected properties of so-obtained correlations, such as no-signaling, indeed hold, and show how to use the framework to derive Bell-inequality-type constraints on these correlations that can be robustly violated by tripartite quantum systems. In particular, we use this framework to rederive a version of one such constraint previously described in a paper of Chao and Reichardt [``Test to separate quantum theory from non-signaling theories,'' arXiv:1706.02008 (2017)].
\end{abstract}

\section{Introduction}

Quantum nonlocality is a phenomenon in which spatially separated observers make measurements the outcomes of which cannot be fully explained by preexisting characteristics of the measured systems. Quantum nonlocality has been demonstrated in experiments \cite{hensen:2015,shalm:2015,giustina:2015,rosenfeld:2017} under careful conditions to rule out any mundane alternative explanations for the observed effects. These experiments are sometimes referred to as ``loophole free." To date, all loophole-free experiments have demonstrated the phenomenon of quantum nonlocality with \textit{two} spatially separated observing parties, and have thus exhibited a form of \textit{bipartite nonlocality.} Notably, this phenomenon does not allow for the sending of signals between the two parties, and so we call it ``bipartite nonsignaling nonlocality.''

Quantum mechanics also predicts the existence of systems that that can be measured by \textit{three} spatially separated observing parties and exhibit quantum nonlocality. This should naturally entail something beyond trivial scenarios where, for instance, two of the three parties observe bipartite nonsignaling nonlocality while the third party's measured statistics are not correlated in any way with the observations of the first two parties. Indeed, definitions of so-called genuine multipartite nonlocality exist \cite{svetlichny:1987,bancal:2013,gallego:2012,dutta:2020}. These definitions take the perspective that the statistics of the three parties should be considered genuinely tripartite nonlocal if they cannot be decomposed into convex combinations of constituent probability distributions that each allow the exchange of signals between (only) two parties. As described in Refs.~\cite{svetlichny:1987,bancal:2013,gallego:2012,dutta:2020}, there are quantum states and measurements that can demonstrate such versions of genuine tripartite quantum nonlocality.

Here we examine the question of tripartite nonlocality from a slightly different perspective. Accepting that bipartite nonsignaling nonlocal systems exist as a physical phenomenon, we argue that if the observed statistics of a tripartite experiment can be explained by an underlying collection of bipartite nonsignaling nonlocal subsystems, then these tripartite statistics should not be taken as evidence of a categorically new phenomenon beyond the sort of bipartite nonsignaling nonlocality observed in Refs.~\cite{hensen:2015,shalm:2015,giustina:2015,rosenfeld:2017}. The question is then whether there exist tripartite correlations that cannot be explained by a collection of underlying systems exhibiting (only) bipartite nonlocal nonsignaling behavior. This approach differs from the definitions of Refs.~\cite{svetlichny:1987,bancal:2013,gallego:2012,dutta:2020} in prohibiting components of the underlying unobserved subsystems from sending superluminal signals. Interestingly, the resulting class of behaviors turns out to not be strictly smaller, as the underlying subsystems are allowed to interact in a cascaded manner that is not replicable within the frameworks of Refs.~\cite{svetlichny:1987,bancal:2013,gallego:2012,dutta:2020}. The scenario of underlying systems of bipartite nonlocal nonsignalling systems interacting sequentially has been previously described as nonlocal ``boxes" connected with ``wirings" \cite{short:2006,lang:2014}.

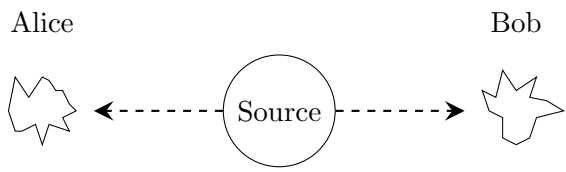
\begin{figure}[p]
    \caption{\textbf{A tripartite system of underlying bipartite nonsignaling nonlocal subystems}}\label{f:overview}
    \centering
    \begin{subfigure}{0.48\textwidth}
        \centering
        \vspace{-.3cm}\begin{tikzpicture}[scale=.9]

                \node[draw, circle] at (0,0) (src) {\small Source};
                
\node at (-3,0) (Alice) {};
\node at (3,0) (Bob) {};

\def\ax{-3.5}    
\def\ay{0}
   
\def\bx{3.5}    
\def\by{0}

\draw (\bx+.7,\by) -- (\bx +.3 ,\by -.1) -- (\bx+.2 ,\by -.4) -- (\bx,\by-.5)  -- (\bx-.2 ,\by-.4 )  -- ( \bx-.2,\by -.1)  -- (\bx-.4 ,\by )  -- ( \bx-.5,\by + .3 )-- ( \bx-.3,\by + .2 ) -- ( \bx-.2,\by + .6 ) -- ( \bx,\by + .3 ) -- ( \bx+.3,\by + .5 )-- ( \bx+.2,\by + .2 )-- ( \bx+.45,\by + .15 )-- cycle;

\draw (\ax+.5,\ay) -- (\ax +.3,\ay-.1 ) -- (\ax +.4,\ay-.3 ) -- (\ax+.1 ,\ay -.2) -- (\ax,\ay-.5)  -- (\ax -.1,\ay -.2) -- (\ax-.3 ,\ay-.3 )  -- (\ax-.4 ,\ay -.3)  -- (\ax-.5 ,\ay )  -- (\ax -.4 ,\ay+.5 )  -- (\ax -.2 ,\ay +.2 ) -- ( \ax,\ay + .5 ) -- ( \ax+.1,\ay + .45 )  -- ( \ax+.2,\ay + .3 ) -- ( \ax+.3,\ay + .3 ) -- ( \ax+.4,\ay +.1 ) -- cycle;

\node at (\ax,1)[above]  {\small Alice};
\node at (\bx,1) [above] {\small Bob};

\draw[decoration={markings,mark=at position .95 with {\arrow[scale=1.8,>=stealth]{>}}},postaction={decorate},shorten >=5, thick,dashed] (src) -- (Alice);
\draw[decoration={markings,mark=at position .95 with {\arrow[scale=1.8,>=stealth]{>}}},postaction={decorate},shorten >=5, thick,dashed] (src) -- (Bob);

\end{tikzpicture}

\vspace{.35cm}

    \caption{\textbf{Bipartite scenario.} Two parties, Alice and Bob, measure particles emitted from a source. If the particles are entangled, the experiment can display the bipartite nonsignaling nonlocality phenomenon.}
    \end{subfigure}
    \hfill
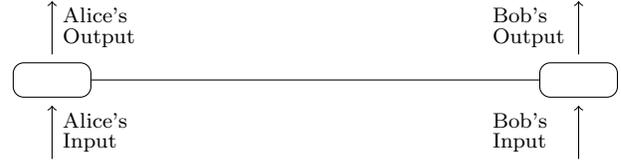
    \begin{subfigure}[t]{0.48\textwidth}
        \centering
   \begin{tikzpicture}

\def\ax{-3.5}    
\def\ay{0}
   
\def\bx{3.5}    
\def\by{0}

\node at (\ax,0)[draw,rounded corners] (Alice) {\textcolor{white}{area}};
\node at (\bx,0)[draw,rounded corners] (Bob) {\textcolor{white}{area}};
\draw  (Alice) -- (Bob) ;

\node at (\ax,-1.3) (Ainp) {};
\node at (\bx,-1.3) (Binp) {};
\node at (\ax,1.3) (Aout) {};
\node at (\bx,1.3) (Bout) {};

 \draw[shorten >=3,shorten <=3,->] (Ainp) -- (Alice)node[pos = 0.5, right, text width = 1.5cm,execute at begin node=\setlength{\baselineskip}{1.5ex}] {\scriptsize Alice's  Input};
 \draw[shorten >=3,shorten <=3,->] (Alice) -- (Aout)node[pos = 0.5, right, text width = 1.5cm,execute at begin node=\setlength{\baselineskip}{1.5ex}] {\scriptsize Alice's  Output};
 
  \draw[shorten >=3,shorten <=3,->] (Binp) -- (Bob)node[pos = 0.5, left, text width = 1cm,execute at begin node=\setlength{\baselineskip}{1.5ex}] {\scriptsize Bob's  Input};
  \draw[shorten >=3,shorten <=3,->] (Bob) -- (Bout)node[pos = 0.5, left, text width = 1cm,execute at begin node=\setlength{\baselineskip}{1.5ex}] {\scriptsize Bob's  Output};
 
  
\end{tikzpicture}

        \caption{\textbf{Schematic depiction.} A pair of particles displaying bipartite nonsignaling effects can be considered as a single system represented as two boxes connected by a line. Input arrows represent measurement settings that Alice and Bob choose; output arrows represent observed measurement outcomes.}  \label{f:bpart}
    \end{subfigure}
    
    
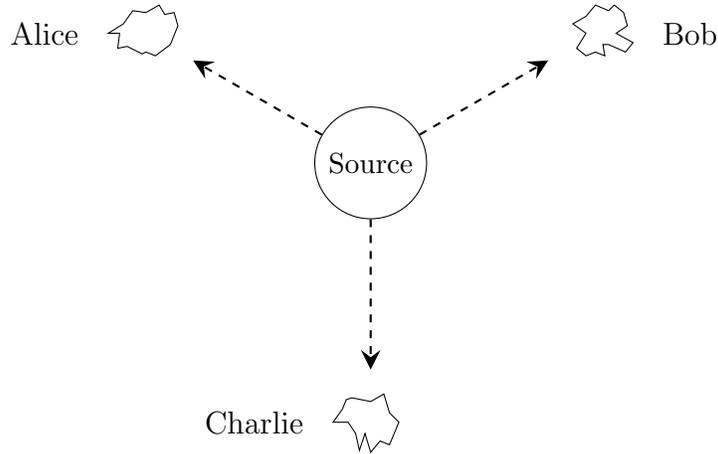
\vspace{.5cm}
    
        \begin{subfigure}[t]{\textwidth}
        \centering
        \begin{tikzpicture}[scale=.25]

\def\eqtri{12}

\node[draw, circle] at (0,0) (src) {\small Source};
\node at (0,-1*\eqtri) (ch) {} ;
\node at (.866*\eqtri,.5*\eqtri) (bo) {};
\node at (-.866*\eqtri,.5*\eqtri) (al) {} ;

\def\ax{-.866*\eqtri*1.15}    
\def\ay{.5*\eqtri*1.15}
   
\def\bx{.866*\eqtri*1.15}    
\def\by{.5*\eqtri*1.15}
   
\def\cx{0}    
\def\cy{-\eqtri*1.15}

\draw[decoration={markings,mark=at position .95 with {\arrow[scale=1.8,>=stealth]{>}}},postaction={decorate},shorten >=5, thick,dashed] (src) -- (ch);
\draw[decoration={markings,mark=at position .95 with {\arrow[scale=1.8,>=stealth]{>}}},postaction={decorate},shorten >=5, thick,dashed] (src) -- (bo);
\draw[decoration={markings,mark=at position .95 with {\arrow[scale=1.8,>=stealth]{>}}},postaction={decorate},shorten >=5, thick,dashed] (src) -- (al);

\draw (-2+\ax,0+\ay) -- (-1.4+\ax,0+\ay) -- (-1.5+\ax,-.8+\ay) -- (-1+\ax,-.7+\ay) -- ( -.2+\ax,-1.1+\ay ) -- ( 0+\ax, -1+\ay) -- ( .5+\ax,-1.2+\ay) -- ( 1.3+\ax,-.6 +\ay) -- ( 1.7+\ax,.4 +\ay) -- ( 1.5+\ax,1.1 +\ay)
 -- ( 1+\ax,1+\ay ) -- ( .7+\ax, 1.5+\ay) -- ( 0+\ax, 1.1+\ay) -- ( -.7+\ax, 1.2+\ay) -- ( -1.2+\ax,.5+\ay ) -- cycle;

\draw (-.5+\bx,0+\by) -- (-1+\bx,-.8+\by)  -- ( -.5+\bx,-1.2+\by ) -- ( 0+\bx, -1+\by) -- ( .5+\bx,-1.2+\by) -- ( .4+\bx,-.6 +\by) -- ( .7+\bx,-.6 +\by)  -- ( 1.6+\bx,-1 +\by)-- ( 2+\bx,-.5 +\by)-- ( 1.1+\bx,0 +\by)-- ( 1.7+\bx,.4 +\by) -- ( 1.5+\bx,1.1 +\by)
 -- ( 1+\bx,1.5+\by ) -- ( .7+\bx, 1.2+\by) -- ( 0+\bx, 1.5+\by) -- ( -.5+\bx, .9+\by) -- ( -1.2+\bx,.5+\by ) -- cycle;

\draw (-2+\cx,0+\cy) -- (-1.2+\cx,0+\cy) -- (-.8+\cx,-.6+\cy) -- ( -.6+\cx,-1.5+\cy ) -- ( -.3+\cx, -.6+\cy) -- ( 0+\cx,-1.6+\cy) -- ( .5+\cx,-1 +\cy) -- ( 1+\cx,-1.2 +\cy) -- ( 1.5+\cx,0 +\cy)
 -- ( 1+\cx,.5+\cy ) -- ( .7+\cx, 1.5+\cy) -- ( 0+\cx, 1.1+\cy) -- ( -1+\cx, 1.3+\cy) -- ( -1.3+\cx, 1.2+\cy) -- ( -1.5+\cx,.7+\cy ) -- cycle;

 \node at (\ax-3,\ay)[left]  {Alice};
\node at (\bx+3,\by) [right] {Bob};
\node at (\cx-3,\cy) [left] {Charlie};

\end{tikzpicture}
    \caption{\textbf{Tripartite scenario.} Here, three parties Alice, Bob and Charlie each receive a particle from the source.}
    \end{subfigure}
    \hfill
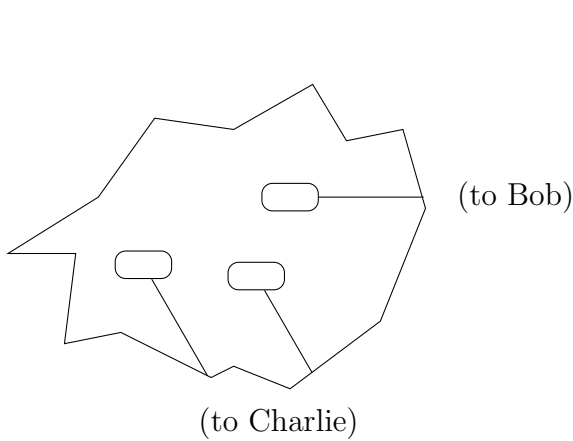
    \begin{subfigure}[t]{0.48\textwidth}
        \centering
        \begin{tikzpicture}[scale=1.5]

\draw (-2,0) -- (-1.4,0) -- (-1.5,-.8) -- (-1,-.7) -- ( -.2,-1.1 ) -- ( 0, -1) -- ( .5,-1.2) -- ( 1.3,-.6 ) -- ( 1.7,.4 ) -- ( 1.5,1.1 )
 -- ( 1,1 ) -- ( .7, 1.5) -- ( 0, 1.1) -- ( -.7, 1.2) -- ( -1.2,.5 ) -- cycle;

 \def\edone{1.1}
 \def\edtwo{1.28}
 \def\edthree{1.25}
 
 \def\ax{.2}
 \def\ay{-.2}
 
 \def\bx{.5}
 \def\by{.5}
 
 \def\cx{-.8}
 \def\cy{-.1}
 
 \node at (\ax,\ay)[draw,rounded corners] (boxone) {\tiny \textcolor{white}{area}};
 \node at (\ax+\edone*.5,\ay-.866*\edone) (Eboxone) {};
 \draw  (boxone) -- (Eboxone);
 
  \node at (\bx,\by)[draw,rounded corners] (boxtwo) {\tiny \textcolor{white}{area}};
 \node at (\bx+\edtwo,\by) (Eboxtwo) {};
 \draw (boxtwo) -- (Eboxtwo);
 
 \node at (\cx,\cy)[draw,rounded corners] (boxthree) {\tiny \textcolor{white}{area}};
 \node at (\cx+\edthree*.5,\cy-.866*\edthree) (Eboxthree) {};
 \draw  (boxthree) -- (Eboxthree);
 
 \node at  (.4,-1.5) {(to Charlie)};
 \node at  (2.5,.5) {(to Bob)};


\end{tikzpicture}
        \caption{\textbf{Detail at Alice.} Alice's particle could be an ensemble of bipartite subsystems like those in (b), each shared with a single other party. In the figure, Alice's particle contains one bipartite system connected with Bob and two bipartite systems connected with Charlie.}  \label{f:dpart}
    \end{subfigure}
\hfill
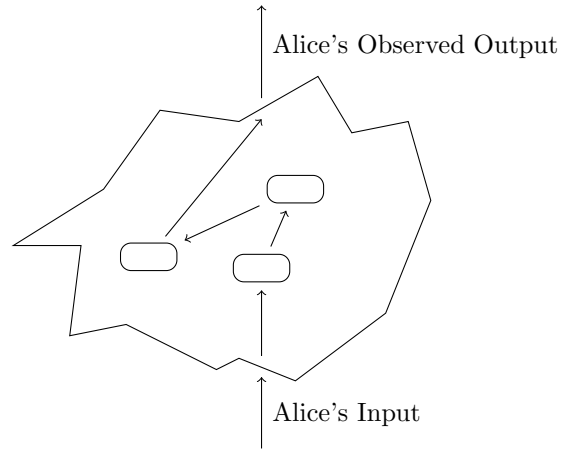
    \begin{subfigure}[t]{.48 \textwidth}
    \centering
            \begin{tikzpicture}[scale=1.5]

\draw (-2,0) -- (-1.4,0) -- (-1.5,-.8) -- (-1,-.7) -- ( -.2,-1.1 ) -- ( 0, -1) -- ( .5,-1.2) -- ( 1.3,-.6 ) -- ( 1.7,.4 ) -- ( 1.5,1.1 )
 -- ( 1,1 ) -- ( .7, 1.5) -- ( 0, 1.1) -- ( -.7, 1.2) -- ( -1.2,.5 ) -- cycle;

 \def\edone{1.18}
 \def\edtwo{1.28}
 \def\edthree{1.92}
 
 \def\ax{.2}
 \def\ay{-.2}
 
 \def\bx{.5}
 \def\by{.5}
 
 \def\cx{-.8}
 \def\cy{-.1}
 
 \node at (\ax,\ay)[draw,rounded corners] (boxone) {\tiny \textcolor{white}{area}};
 \node at (\ax+\edone*.866,\ay-.5*\edone) (Eboxone) {};
 
  \node at (\bx,\by)[draw,rounded corners] (boxtwo) {\tiny \textcolor{white}{area}};
 \node at (\bx+\edtwo,\by) (Eboxtwo) {};
 
 \node at (\cx,\cy)[draw,rounded corners] (boxthree) {\tiny \textcolor{white}{area}};
 \node at (\cx+\edthree*.866,\cy-.5*\edthree) (Eboxthree) {};

 \draw[shorten >=5,->] ( 0.2, -1.8) --( 0.2, -1.05)node[pos = 0.4, right,execute at begin node=\setlength{\baselineskip}{1.5ex}] {\footnotesize Alice's Input};
 \draw[shorten >=3,shorten <=3,->] ( 0.2, -1.05) --(boxone);
 \draw[shorten >=3,shorten <=3,->]  (boxone)--(boxtwo);
 \draw[shorten >=3,shorten <=3,->] (boxtwo)--(boxthree);
 \draw[shorten >=0,shorten <=3,->] (boxthree)--(0.2,1.12);
  \draw[shorten >=3,shorten <=8,->] ( 0.2, 1.12) --( 0.2, 2.2)node[pos = 0.6, right,execute at begin node=\setlength{\baselineskip}{1.5ex}] {\footnotesize Alice's Observed Output};

\end{tikzpicture}

    \caption{\textbf{Alice's measurement.} When Alice performs a measurement and provides an input, the bipartite subsystems are measured in some order with inputs to later sub-measurements possibly depending on outcomes of earlier sub-measurements. She does not directly observe the sub-measurements, but her final observed outcome is some function of the sub-measurement outcomes.}  \label{f:epart}
    \end{subfigure}

\end{figure}

To describe our perspective, illustrated in Figure \ref{f:overview}, consider a scenario of three spatially separated observing parties, Alice, Bob, and Charlie, each of whom make a measurement. Suppose what Alice actually measures is an ensemble of subsystems, each one of them a bipartite nonsignaling nonlocal system, some shared with Bob and some with Charlie. Alice's macro-measurement induces these subsystems to be measured in some sequential order, with the outcomes of early sub-measurements possibly affecting the progression of later sub-measurements, and her final observed output is a function of the outcomes of the sub-measurements. Bob and Charlie's observed outputs are obtained in a similar manner from cascaded measurements of their respective halves of the bipartite nonsignaling nonlocal subsystems shared between pairs of parties. Now, if the statistics of the three-party experiment can be explained by a model of this sort, then the experiment can be said to be consistent with the existence of (only) the phenomenon of bipartite nonsignaling nonlocal systems. Conversely, if the measured statistics are inconsistent with any such model, then a new form of tripartite nonlocal behavior has been observed. Interestingly, some correlations that display the forms of genuine tripartite nonlocality defined in \cite{svetlichny:1987,bancal:2013,gallego:2012,dutta:2020}, such as the distributions numbered 44-46 in \cite{pironio:2011}, can be replicated by bipartite-ensemble systems, as shown in Section III.C of \cite{barrett:2005}.

The mathematical problem of ruling out bipartite ensemble models for a given tripartite behavior was first examined \cite{barrettpironio:2005,scarani:9.2.2} in regards to the question of whether a certain canonical bipartite nonsignaling nonlocal correlation, the so-called PR Box (named for Popescu and Rohrlich \cite{PRBOX}), can be considered as a unit of nonlocality insofar as it is able to simulate other nonlocal correlations. These initial results showed that certain multi-party correlations cannot be reproduced exactly by systems of PR boxes \cite{barrettpironio:2005,scarani:9.2.2}, but this did not rule out the possibility of $\epsilon$-close simulations. In contrast, a more recent claim of Chao and Reichardt \cite{chao:2017} describes a robust separation between a quantum-achievable probability distribution and all bipartite-simulable correlations. Ref.~\cite{chao:2017} also studies $n$-party generalizations of this problem; other recent relevant results, some of which are noise-robust, include \cite{anshu:2013,mathieu:2018,anshu:2020}. 

The argument of Ref.~\cite{chao:2017}, applied to the 3-party case, introduces random variables that ``parametrize the randomness" of the underlying bipartite nonsignaling nonlocal subsystems, then uses functional dependencies between the introduced random variables and the outcomes of the bipartite subsystems to derive a constraint -- a Bell-like inequality -- on the joint probability distribution of observed outcomes. The constraint can be robustly violated by an appropriate quantum system. However, the mathematical setting for working with the parametrizing random variables is imprecise, obscuring the justification for some of the claimed functional dependencies between the introduced random variables and the existing ones. 

The goal of this paper is to provide a rigorous mathematical framework for proving the Chao and Reichardt constraint in the tripartite case. Rather than introducing new random variables, we work directly with the existing output variables of the bipartite nonsignaling nonlocal subsystems and derive functional relationships between them, leading to arguably a conceptual simplification over the approach of Ref.~\cite{chao:2017}. Our approach leverages a key result of Forster and Wolf \cite{forster:2011}: PR boxes can simulate general bipartite nonsignaling nonlocal correlations to arbitrary accuracy. (This generalizes an earlier result showing simulation is possible for bipartite correlations with any number of measurement settings but restricted to binary outputs \cite{jones:2005}.) The result of Ref.~\cite{forster:2011} implies that one can consider only networks where all the underlying subsystems are PR boxes without any loss of generality.

We use this framework to clarify the natural conditions that imply the existence of a unique global joint distribution for the outputs of the ensembles of bipartite nonsignaling nonlocal subsystems that is consistent with any temporal ordering in which the parties measure the subsystems. We also show that this global distribution necessarily obeys desired no-signaling conditions which are required to derive the the main result. (That a network of bipartite nonsignaling correlations is nonsignaling in the aggregate might be unsurprising, but it is useful to illustrate precisely how this follows from the assumptions.) Finally, we make a minor modification to adapt the Chao and Reichardt constraint to a form more applicable to an experiment of three space-like separated parties that measure systems near-simultaneously with randomized settings choices. The modification allows us to avoid the ``external verifier'' paradigm of Ref.~\cite{chao:2017}, where verifier-chosen measurement settings lead to setting probabilities that are not independent between the observing parties. We can instead employ a uniform setting probability distribution that allows an experimental scenario where each of the space-like separated parties chooses measurement settings locally based on a random process, independently of the other parties, as is standard in the experimental setups of Refs.~\cite{hensen:2015,shalm:2015,giustina:2015,rosenfeld:2017}.


In Section \ref{s:groundwork}, we develop properties of the joint distributions of collections of bipartite nonsignaling nonlocal subsystems shared among three parties, and show that the joint distribution is determined uniquely by a few simple principles. Using these results, we can follow the outline of the argument in Ref.~\cite{chao:2017} to prove in Section \ref{s:CR} the constraint (a Bell-like inequality) for the three-party scenario that must be satisfied by any experimental behavior induced by an underlying ensemble of bipartite nonsignaling nonlocal subsystems, which can be violated by quantum mechanics. Section \ref{s:conclusion} contains concluding remarks. Appendices provide an example of how a candidate joint distribution for a collection of bipartite systems can display unphysical characteristics if proper care is not taken in its construction, and a proof for a bound on a sum of probabilities used in the main text.

\section{Determining Joint Probability Distributions for interconnected PR Boxes}\label{s:groundwork}

Our experimental scenario is the (3,2,2) setting where three parties named Alice, Bob and Charlie each choose between two measurement settings and observe one of two possible outcomes. We denote the measurement choices of Alice, Bob and Charlie with random variables $X$, $Y$, and $Z$ taking values in the respective sets $\{\mathsf{a}, \mathsf{a}\textnormal{\textquotesingle}\}$, $\{\mathsf{b}, \mathsf{b}\textnormal{\textquotesingle}\}$, and $\{\mathsf{c}, \mathsf{c}\textnormal{\textquotesingle}\}$, and we represent the observed outcomes with random variables $A$, $B$, and $C$ all taking values in the set $\{0,\text{+}\}$. All settings configurations will be equiprobable, so $P(X,Y,Z)=1/8$. In each round, each party always records an output, even if the output does not always factor into the final statistical analysis. 

We ask what sort of observable distributions $\pp(A,B,C|X,Y,Z)$ are possible if the underlying system being measured consists of unobserved ensembles of bipartite nonsignaling nonlocal subsystems shared between pairs of parties, possibly supplemented with shared local randomness, and the observable outcomes $A$, $B$ and $C$ are local functions of the outputs of these subsystems and the local settings. To address the question, we must first formalize the notion of an underlying ensemble of bipartite nonsignaling nonlocal subsystems.

\subsection{Networks of PR Boxes}

We consider the possible characteristics of the joint distribution of a collection of sub-systems obeying the PR box behavior. The PR box is a specific bipartite nonsignaling nonlocal behavior where each of the two parties sharing it has a binary input and a binary output. Importantly, any bipartite nonsignaling nonlocal distribution with a finite number of inputs and outputs can be simulated to arbitrary precision with collections of PR boxes and shared local randomness \cite{forster:2011}, so results obtained for ensembles of PR boxes will apply generally: any system involving non-PR box bipartite subsystems can be approximated to any desired accuracy by replacing each non-PR box subsystem with a collection of PR boxes that approximates its behavior. While some of these approximations require shared local randomness, for our discussion we neglect the shared local randomness and only consider collections of PR boxes. This is acceptable because the Bell function we will upper bound for PR-box-network-induced distributions is linear in the value of any shared local randomness and so upper bounds on this function cannot be circumvented by taking convex mixtures of distributions with different values of shared local randomness.

We write the conditional probability distribution of an individual PR Box as $\pp(a,b|x,y)$, where $a$ and $b$ are the respective outputs for the two parties sharing the box and $x$ and $y$ are their respective inputs. The conditional probabilities are shown in the table and corresponding formula below, where the symbol $\oplus$ denotes addition modulo 2:

\medskip

\medskip

\begin{minipage}{0.45\textwidth}
     \begin{tabular}{ |cr|cccc| }
     \multicolumn{6}{c}{Table of $\pp(a,b|x,y)$ values}\\
     \hline
& \multicolumn{1}{r}{} &\multicolumn{4}{|c|}{$a,b$}
 \\
& \multicolumn{1}{r}{}
  &  \multicolumn{1}{|c}{$0,0$}
 &  \multicolumn{1}{c}{$0,1$}
 &  \multicolumn{1}{c}{$1,0$} 
   &  \multicolumn{1}{c|}{$1,1$}
\\
\cline{1-6}
  & $0,0$&   1/2  &  0  &  0  & 1/2 \\
$x,y$  & $0,1$&   1/2  &  0  &  0  & 1/2 \\
  & $1,0$&   1/2 &  0  &  0  & 1/2 \\
  & $1,1$&   0  &  1/2  &  1/2  & 0 \\
  \hline
 \end{tabular}
\end{minipage}
\begin{minipage}{0.51\textwidth}
    \begin{equation*}
    \text{For all } a,b,x,y \in \{0,1\},
    \end{equation*}
    \begin{equation}\label{e:PRprobs}
    \pp(a,b|x,y)=\begin{cases} 
      1/2 & \text{if }a\oplus b = xy \\
      0 & \text{otherwise} 
   \end{cases}
    \end{equation}
\end{minipage}

\medskip

\medskip

\noindent From the above formula we have the following properties, which also hold if $a$ and $b$ are switched:
\begin{eqnarray}
&&\text{The marginal of $a$, $\pp(a|xy)$, is uniform on $\{0,1\}$ independent of inputs $x,y$}\label{e:unifmarg}\\
&&\text{If $a$, $x$, and $y$ are given, $b$ is completely determined: $\pp(b|axy)=\delta_{b=a\oplus xy}$}\label{e:determined}
\end{eqnarray}

\def\stc{1.5}
\def\aloc{2}
\def\psep{6.5}
\def\minsize{25}
\begin{figure}\caption{\textbf{Schematic Depiction of PR Boxes}}
\centering
    \begin{subfigure}{1\textwidth}
  \centering 
\begin{tikzpicture}
        \node[draw,minimum size=60,text width = 1.7 cm, text centered] at (0,2.5) (Aout) {Alice's \ output $a$: \ 0 or 1};
        \node[draw,minimum size=60,text width = 1.7 cm, text centered] at (8,2.5) (Bout) {Bob's \ output $b$: \ 0 or 1};
        \node[text width = 1.5 cm, text centered] at (0,0) (Ain) {Alice's input $x$: 0 or 1};
        \node[text width = 1.5 cm, text centered] at (8,0) (Bin) {Bob's input $y$: 0 or 1};
        
	\draw[decoration={markings,mark=at position .9 with {\arrow[scale=1.5,>=stealth]{>}}},postaction={decorate},shorten >=3] (Ain) -- (Aout);
	\draw[decoration={markings,mark=at position .9 with {\arrow[scale=1.5,>=stealth]{>}}},postaction={decorate},shorten >=3] (Bin) -- (Bout);
	\draw[-,dashed] (Aout) -- (Bout);
        
\end{tikzpicture}
\vspace{5mm}
\end{subfigure}\subcaption{\textbf{A single PR box shared between Alice and Bob.} The dashed line, indicating that Alice's and Bob's outputs are coming from the same PR box, will usually be suppressed. We have slightly modified the representation of the outputs compared to Figure \ref{f:overview}, now displaying them as being contained in squares, as opposed to being carried on arrows exiting the system. This is to avoid potential ambiguity between inputs and outputs in future diagrams where multiple PR boxes are measured in sequence.}\label{f:onebox}
\vspace{8mm}
\begin{subfigure}{1\textwidth}
\centering
\begin{tikzpicture}

        \node[draw,minimum size = \minsize] at (\aloc,-1) (a1) {};
        \node[draw,minimum size = \minsize] at (\aloc,-1-\stc) (a2) {};
        \node[draw,minimum size = \minsize] at (\aloc,-1-2*\stc) (a3) {${\text{\tiny{Out-}}\atop\text{\tiny{put}}}$};
        \node[draw,minimum size = \minsize] at (\aloc+\stc,-1) (a4){};
        \node[draw,minimum size = \minsize] at (\aloc+\stc,-1-\stc) (a5){};
        \node[draw,minimum size = \minsize] at (\aloc+\stc,-1-2*\stc) (a6){};
        \node[circle,draw,minimum size = \minsize] at (\aloc+0.5*\stc,-1-3*\stc) (a7){$X$};
        \node at (\aloc+0.5*\stc,-1-3.5*\stc) (a8){Alice};

        \node[draw,minimum size = \minsize] at (\aloc-\psep,-1) (b1) {};
        \node[draw,minimum size = \minsize] at (\aloc-\psep,-1-\stc) (b2) {};
        \node[draw,minimum size = \minsize] at (\aloc-\psep,-1-2*\stc) (b3) {};
        \node[draw,minimum size = \minsize] at (\aloc-\psep+\stc,-1) (b4){};
        \node[draw,minimum size = \minsize] at (\aloc-\psep+\stc,-1-\stc) (b5){};
        \node[draw,minimum size = \minsize] at (\aloc-\psep+\stc,-1-2*\stc) (b6){${\text{\tiny{Out-}}\atop\text{\tiny{put}}}$};
        \node[circle,draw,minimum size = \minsize] at (\aloc-\psep+0.5*\stc,-1-3*\stc) (b7){$Y$};
        \node at (\aloc-\psep+0.5*\stc,-1-3.5*\stc) (b8){Bob};

        \node[draw,minimum size = \minsize] at (\aloc+\psep,-1) (c1) {};
        \node[draw,minimum size = \minsize] at (\aloc+\psep,-1-\stc) (c2) {};
        \node[draw,minimum size = \minsize] at (\aloc+\psep,-1-2*\stc) (c3) {};
        \node[draw,minimum size = \minsize] at (\aloc+\psep+\stc,-1) (c4){};
        \node[draw,minimum size = \minsize] at (\aloc+\psep+\stc,-1-\stc) (c5){};
        \node[draw,minimum size = \minsize] at (\aloc+\psep+\stc,-1-2*\stc) (c6){};
        \node[circle,draw,minimum size = \minsize] at (\aloc+\psep+0.5*\stc,-1-3*\stc) (c7){$Z$};
        \node at (\aloc+\psep+0.5*\stc,-1-3.5*\stc) (c8){Charlie};
        
                \draw[shorten >=3,shorten <=3,->] (a7) -- node[left] {\tiny{Input}} (a3);
                \draw[shorten >=3,shorten <=3,->] (b7) -- node[right] {\tiny{Input}} (b6);
        
               \draw[-,dashed] (b6) --node [above,midway] {\small a PR box shared} node [below,midway] {\small by Bob and Alice}(a3);
               \draw[-,dashed] (c2) --node [above,midway] {\small a PR box shared} node [below,midway] {\small by Alice and Charlie}(a5);

\end{tikzpicture}
\vspace{5mm}
\caption{\textbf{Multiple PR boxes possessed by pairs of parties in a tripartite scenario.} Alice and Bob share three PR boxes, as do Alice and Charlie, as do Bob and Charlie with Charlie's right column of output boxes connecting to Bob's left column. The circles at the bottom represent the measurement settings that Alice, Bob, and Charlie apply to their overall systems when the experiment is performed. The arrows at the bottom of the figure imply that the choice of input to Bob's half of a PR box subsystem shared with Alice can depend on his setting $Y$, and the choice of input to Alice's half can depend on $X$.}\label{f:manyprs} 

\end{subfigure}

\end{figure}
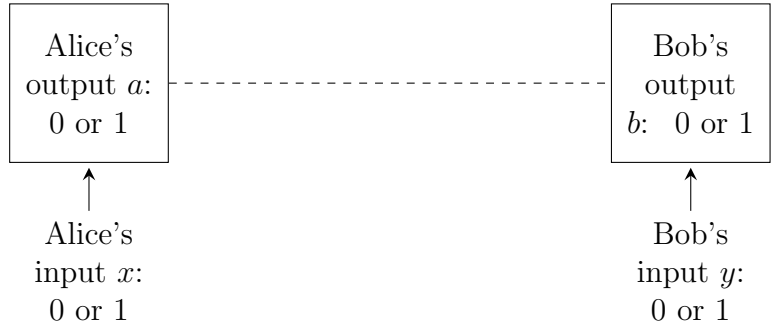

It is useful to depict single PR boxes schematically as in Figure \ref{f:onebox} with arrows representing the inputs and squares containing the outputs. Figure \ref{f:manyprs} uses this scheme to illustrate a complicated overall system where Alice, Bob, and Charlie share multiple PR boxes pairwise. It is important to remember that the network in Figure \ref{f:manyprs} is not directly observed by the parties when they measure the system. Alice, for example, only sees a final output $A\in \{\text{+},0\}$ after supplying her measurement setting $X$. But in our model, we posit that the macro-measurement process induces the PR box subsystems to be queried according to a set pattern which can depend on the setting $X$; the final observed output $A$ is then a function of the PR box outputs and $X$. It is convenient to refer to the set pattern for querying the PR boxes as Alice's \textit{strategy}, even though she does not observe the PR boxes and does not know how they are being used.

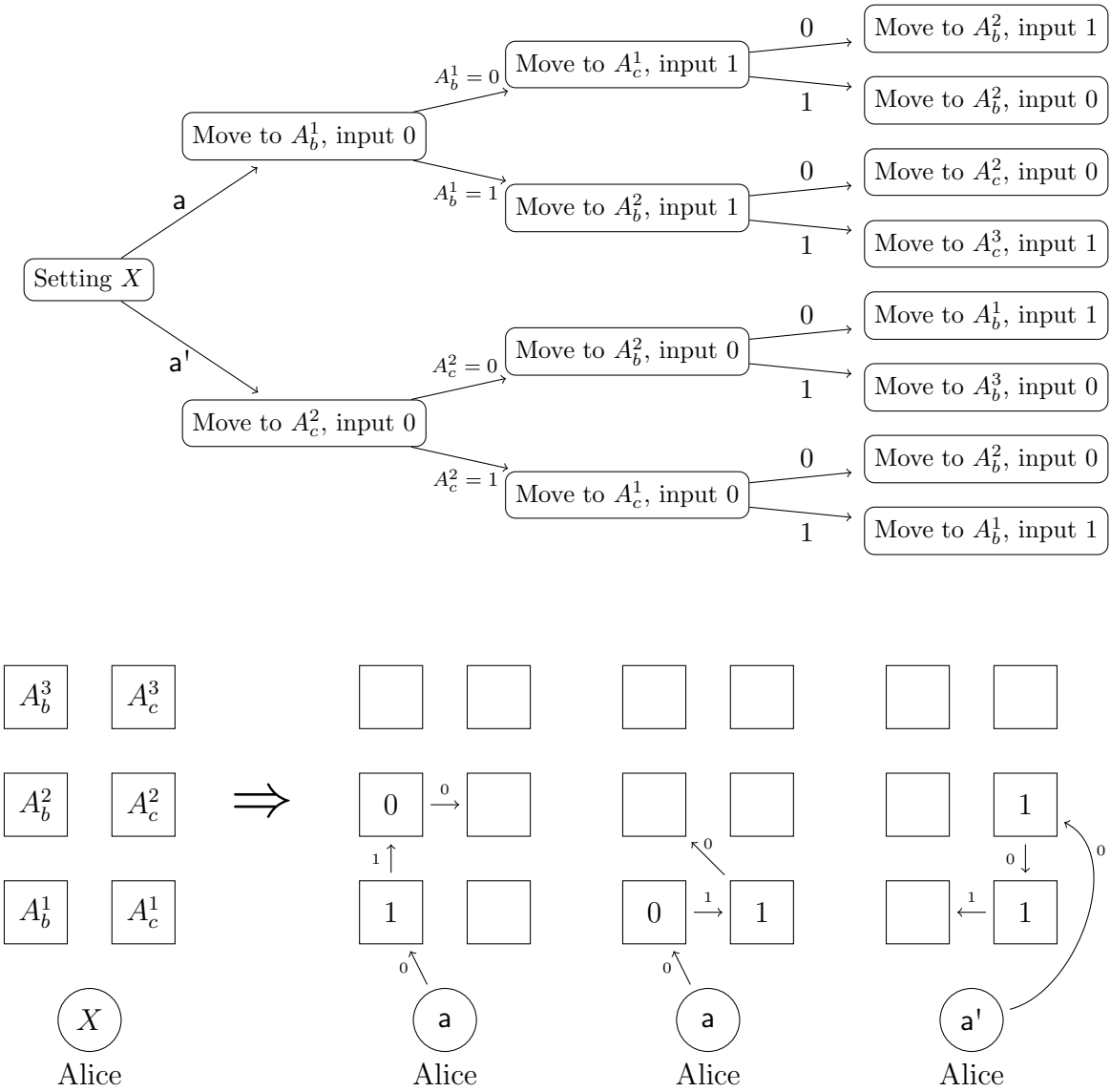
\begin{figure}[p!]\caption{\textbf{A decision tree.} The following example is for a scenario where Alice shares three PR boxes with Bob and three PR boxes with Charlie. Only the first three branch points of Alice's decision tree are displayed; a full decision tree for this scenario would continue to the right for four more columns. The bottom of the figure shows examples of a few possible paths Alice could take following this decision tree for various choices of her setting and different observed outputs of the PR boxes.}\label{f:strategy}
\centering
\tikzstyle{bag} = [circle,draw]
\tikzstyle{square} = [rectangle,draw]
\tikzstyle{level 1}=[level distance=3cm, sibling distance=4cm]
\tikzstyle{level 2}=[level distance=4.5cm, sibling distance=2cm]
\tikzstyle{level 3}=[level distance=5cm, sibling distance=1cm]
\begin{tikzpicture}[grow=right,->,shorten >=5]
\node[rectangle,rounded corners,draw] {\footnotesize Setting $X$}
	child {
	node[rectangle,rounded corners,draw] {\footnotesize Move to $A_c^2$, input 0}      
		child {
            	node[rectangle,rounded corners,draw] {\footnotesize Move to $A_c^1$, input 0}   
			child {
	         	node[rectangle,rounded corners,draw] {\footnotesize Move to $A_b^1$, input 1}         
	                 edge from parent 
	                 node[below]  {\small$1$}
	           	}
	           	child {
	            	node[rectangle,rounded corners,draw] {\footnotesize Move to $A_b^2$, input 0}        
	                 edge from parent 
	                	node[above]  {\small$0$}
           		}      
		edge from parent 
		node[below]  {\scriptsize$A_c^2=1$}
		}
        		child {
        		node[rectangle,rounded corners,draw] {\footnotesize Move to $A_b^2$, input 0}   
			child {
	         	node[rectangle,rounded corners,draw] {\footnotesize Move to $A_b^3$, input 0}        
	                 edge from parent 
	                 node[below]  {\small $1$}
	           	}
	           	child {
	            	node[rectangle,rounded corners,draw] {\footnotesize Move to $A_b^1$, input 1}        
	                 edge from parent 
	                	node[above]  {\small $0$}
           		}          
        		edge from parent 
        		node[above]  {\scriptsize $A_c^2=0$}
        		}
	edge from parent 
	node[below,pos=0.4]  {$\mathsf{a}\textnormal{\textquotesingle}$}
    	}
	child {
	node[rectangle,rounded corners,draw] {\footnotesize Move to $A_b^1$, input 0}        
		child {
            	node[rectangle,rounded corners,draw] {\footnotesize Move to $A_b^2$, input 1}  
			child {
	         	node[rectangle,rounded corners,draw] {\footnotesize Move to $A_c^3$, input 1}        
	                 edge from parent 
	                 node[below]  {\small $1$}
	           	}
	           	child {
	            	node[rectangle,rounded corners,draw] {\footnotesize Move to $A_c^2$, input 0}        
	                 edge from parent 
	                	node[above]  {\small$0$}
           		}      
		edge from parent 
		node[below]  {\scriptsize $A_b^1=1$}
		}
        		child {
        		node[rectangle,rounded corners,draw] {\footnotesize Move to $A_c^1$, input 1}   
			child {
	         	node[rectangle,rounded corners,draw] {\footnotesize Move to $A_b^2$, input 0}        
	                 edge from parent 
	                 node[below]  {\small$1$}
	           	}
	           	child {
	            	node[rectangle,rounded corners,draw] {\footnotesize Move to $A_b^2$, input 1}        
	                 edge from parent 
	                	node[above]  {\small$0$}
           		}          
        		edge from parent 
        		node[above]  {\scriptsize$A_b^1=0$}
        		}
	edge from parent 
	node[above,pos=0.4]  {$\mathsf{a}$}
    	}
    ;
\end{tikzpicture}

\vspace{1.5cm}

\begin{tikzpicture}
\def\stc{1.5}
\def\aloc{2}
\def\minsize{25}
        \node[draw,minimum size = \minsize] at (\aloc,-1) (AB3) {$A_b^3$};
        \node[draw,minimum size = \minsize] at (\aloc,-1-\stc) (AB2) {$A_b^2$};
        \node[draw,minimum size = \minsize] at (\aloc,-1-2*\stc) (AB1) {$A_b^1$};
        \node[draw,minimum size = \minsize] at (\aloc+\stc,-1) (AC3){$A_c^3$};
        \node[draw,minimum size = \minsize] at (\aloc+\stc,-1-\stc) (AC2){$A_c^2$};
        \node[draw,minimum size = \minsize] at (\aloc+\stc,-1-2*\stc) (AC1){$A_c^1$};
        \node[circle,draw,minimum size = \minsize] at (\aloc+0.5*\stc,-1-3*\stc) (setting){$X$};
        \node at (\aloc+0.5*\stc,-1-3.5*\stc) (blocka8){Alice};
        
        \node at (\aloc+2.1*\stc,-1-\stc) (argh){\huge$\Rightarrow$};
        
\end{tikzpicture}
\hspace{5mm}
\begin{tikzpicture}
\def\stc{1.5}
\def\aloc{2}
\def\minsize{25}
        \node[draw,minimum size = \minsize] at (\aloc,-1) (AB3) {};
        \node[draw,minimum size = \minsize] at (\aloc,-1-\stc) (AB2) {0};
        \node[draw,minimum size = \minsize] at (\aloc,-1-2*\stc) (AB1) {$1$};
        \node[draw,minimum size = \minsize] at (\aloc+\stc,-1) (AC3){};
        \node[draw,minimum size = \minsize] at (\aloc+\stc,-1-\stc) (AC2){};
        \node[draw,minimum size = \minsize] at (\aloc+\stc,-1-2*\stc) (AC1){};
        \node[circle,draw,minimum size = \minsize] at (\aloc+0.5*\stc,-1-3*\stc) (setting){$\mathsf{a}$};
        \node at (\aloc+0.5*\stc,-1-3.5*\stc) (blocka8){Alice};

                \draw[shorten >=3,shorten <=3,->] (setting) -- (AB1)node[pos = 0.5, left, text centered] {{\tiny 0}};
        \draw[shorten >=3,shorten <=3,->] (AB1) -- (AB2) node[pos = 0.5, left, text centered] {{\tiny 1}};
        \draw[shorten >=3,shorten <=3,->] (AB2) -- (AC2)node[pos = 0.5, above, text centered] {{\tiny 0}};

\end{tikzpicture}
\hspace{10mm}
\begin{tikzpicture}
\def\stc{1.5}
\def\aloc{2}
\def\minsize{25}
        \node[draw,minimum size = \minsize] at (\aloc,-1) (AB3) {};
        \node[draw,minimum size = \minsize] at (\aloc,-1-\stc) (AB2) {};
        \node[draw,minimum size = \minsize] at (\aloc,-1-2*\stc) (AB1) {0};
        \node[draw,minimum size = \minsize] at (\aloc+\stc,-1) (AC3){};
        \node[draw,minimum size = \minsize] at (\aloc+\stc,-1-\stc) (AC2){};
        \node[draw,minimum size = \minsize] at (\aloc+\stc,-1-2*\stc) (AC1){1};
        \node[circle,draw,minimum size = \minsize] at (\aloc+0.5*\stc,-1-3*\stc) (setting){$\mathsf{a}$};
        \node at (\aloc+0.5*\stc,-1-3.5*\stc) (blocka8){Alice};

                \draw[shorten >=3,shorten <=3,->] (setting) -- (AB1)node[pos = 0.5, left, text centered] {{\tiny 0}};
        \draw[shorten >=3,shorten <=3,->] (AB1) -- (AC1) node[pos = 0.5, above, text centered] {{\tiny 1}};
        \draw[shorten >=3,shorten <=3,->] (AC1) -- (AB2)node[pos = 0.5, above, text centered] {{\tiny 0}};

\end{tikzpicture}
\hspace{10mm}
\begin{tikzpicture}
\def\stc{1.5}
\def\aloc{2}
\def\minsize{25}
        \node[draw,minimum size = \minsize] at (\aloc,-1) (AB3) {};
        \node[draw,minimum size = \minsize] at (\aloc,-1-\stc) (AB2) {};
        \node[draw,minimum size = \minsize] at (\aloc,-1-2*\stc) (AB1) {};
        \node[draw,minimum size = \minsize] at (\aloc+\stc,-1) (AC3){};
        \node[draw,minimum size = \minsize] at (\aloc+\stc,-1-\stc) (AC2){$1$};
        \node[draw,minimum size = \minsize] at (\aloc+\stc,-1-2*\stc) (AC1){$1$};
        \node[circle,draw,minimum size = \minsize] at (\aloc+0.5*\stc,-1-3*\stc) (setting){$\mathsf{a}\textnormal{\textquotesingle}$};
        \node at (\aloc+0.5*\stc,-1-3.5*\stc) (blocka8){Alice};

        \draw[shorten >=3,shorten <=3,->,bend right=90] (setting) to [out=300,in=260]node[pos = 0.8, right, text centered] {{\tiny 0}} (AC2);
        \draw[shorten >=3,shorten <=3,->] (AC2) -- (AC1) node[pos = 0.5, left, text centered] {{\tiny 0}};
        \draw[shorten >=3,shorten <=3,->] (AC1) -- (AB1)node[pos = 0.5, above, text centered] {{\tiny 1}};

\end{tikzpicture}
\end{figure}

To analyze strategies for networks of PR boxes and how they induce the final outputs $A,B,C$, we introduce random vectors ${\bf P}_q$ to denote the string of outputs observed by party $P$ from the PR boxes shared with party $q$, with $P_q^i$ denoting the $i$th element of ${\bf P}_q$. So for example when Alice shares $n$ PR boxes with Bob and $n$ PR boxes with Charlie, her relevant random variables are as follows:
\vspace{-3mm}
\begin{center}
\begin{tabular}{ ccl }
\multicolumn{3}{c}{\underline{Random Variables for Alice}{\color{white}h}}\vspace{3mm} \\
{\it Variable} &{\it Outcome Space} & {\it Meaning} \\
$X$ & $\{\mathsf{a},\mathsf{a}\textnormal{\textquotesingle}\}$&Measurement setting \\
${\bf A}_b$ &$\{0,1\}^n$& String of all PR box outputs from PR boxes shared with Bob \\
${\bf A}_c$ &$\{0,1\}^n$& String of all PR box outputs from PR boxes shared with Charlie \\
$A_b^i$ &$\{0,1\}$& Output of $i$th PR box shared with Bob \\
$ A_c^i$ &$\{0,1\}$& Output of $i$th PR box shared with Charlie  \\
$A$ &$\{0,\text{+}\}$& Final observed output; a function of $X$, ${\bf A}_c$, and ${\bf A}_b$\\
\end{tabular}
\end{center}
\vspace{3mm}
We now proceed to formalize strategies for each party's usage of PR boxes. We want to allow for the possibility of parties feeding the output from one PR box (or the opposite of the output) as input into another PR box and possibly changing the order in which they use later PR boxes based on early PR box outputs. Such a strategy can be called a {\it wiring} to evoke the connecting of multiple PR boxes together \cite{short:2006,lang:2014}. Following \cite{barrett:2005,forster:2011}, a general strategy consists of a list of step-by-step instructions for a party to follow that determines the order in which they query their PR boxes and the inputs they provide them. This list of instructions can be handily visualized as a decision tree that guides the party through their collection of PR boxes; see Figure \ref{f:strategy} for an example. We assume that a party always uses all of their PR boxes; if we want to consider strategies where some remain unused, we model this by having the party input 0 to all unwanted PR boxes at the end of the decision tree and not using those outputs to determine the final output in $\{0,\text{+}\}$.

\subsection{Principles for assigning a joint probability distribution}

We want to develop some characteristics of the joint probability distribution\linebreak $\pp (\textbf{A}_b, \textbf{A}_c, \textbf{B}_a, \textbf{B}_c, \textbf{C}_a, \textbf{C}_b |X,Y,Z)$ of PR box outcomes in the scenario where each of the three parties follows a Figure-\ref{f:strategy}-style decision tree. Naturally, the distribution $\pp(\cdot)$ will depend on the particular decision trees employed by the parties. The distribution can be constructed mathematically by formalizing the natural condition that as an individual party works through the Figure \ref{f:strategy}-style decision tree, the marginal distribution of the outcome of the next PR box must always be uniform, conditioned on the PR box outcomes that the party has already observed as well as the original measurement setting. We need to formalize such an assumption to discount pathological joint distributions where, for instance, all of the PR boxes of one party always yield the same output; an example of this is given in Appendix \ref{s:sigpr}. Once this assumption is made, the joint distribution can be constructed if we pick an arbitrary ordering of parties -- for instance, Charlie, then Alice, then Bob -- and imagine that Charlie works through his entire tree first, then Alice works through her entire tree, followed by Bob, and assign probabilities according to the following rules: 1) every PR box output that the first party, Charlie, observes is uniformly distributed conditioned on PR boxes he has already observed, then 2) as Alice works through her tree, outcomes of PR boxes shared with Bob are uniform and those shared with Charlie are determined by Charlie's already-recorded output via Eq.~\eqref{e:determined}, and 3) all of Bob's PR box outputs are determined via Eq.~\eqref{e:determined} by Alice's and Charlie's outputs as he works through his decision tree. This procedure induces a unique joint distribution for the PR box outputs. 

A natural question is whether this joint distribution is independent of the choice of the ordering of parties used to construct it. Indeed, in a scenario where all parties are space-like separated during the measurement process, there is no preferred order in which the parties measure their systems. Regarding this matter, Ref.~\cite{barrett:2005} asserts that the no-signaling property of PR boxes assures the possibility of assigning a joint distribution consistent with any time ordering, and we will confirm in Subsection \ref{s:consist} that indeed any ordering of parties leads to the same joint distribution.

Our rule for assigning probabilities for the first party is expressed mathematically in the following manner: Supposing Charlie is the party going first, consider some $C_a^i$. Then for any sequence of PR boxes that can precede a querying of $C_a^i$ on a branch of the decision tree, 
\begin{equation}\label{e:stipulation}
\pp(C_a^i=1|X,Y,Z,\{\text{Outcomes of earlier PR boxes on the branch}\})=1/2.
\end{equation}
This holds for each $C_b^i$ as well. The appearance of all three settings in the conditioner, as opposed to just Charlie's setting $Z$, is a necessary assumption in the task of constructing a joint distribution of all the random variables $\{X,Y,Z,\textbf{C}_a,\textbf{C}_b,\textbf{A}_b,\textbf{A}_c,\textbf{B}_c,\textbf{B}_a\}$. If we don't include settings $X$ and $Y$ in the conditioner, we cannot rule out problematic joint distributions where Charlie's output probabilities depend on Alice and/or Bob's settings. Luckily, assuming the independence of $X$ and $Y$ in the form specified by \eqref{e:stipulation} is natural, and motivated by the no-signaling principle; if Charlie's output probabilities ever deviated from 1/2 dependent on $X$ and/or $Y$, this would permit signaling from Alice and/or Bob to Charlie. It is also worth noting that the expressions in \eqref{e:stipulation} depend on Charlie's strategy (dictating which are the ``earlier PR boxes on the branch"), but are independent of Alice's and Bob's strategies, even as these other parties could be doing any manner of things observing PR boxes from Charlie and feeding these outputs to other PR boxes shared with Charlie and/or each other. 

Once the first party (Charlie) has moved through his decision tree and recorded all PR box outputs, we move on to the decision tree of the second party (Alice). When boxes shared with Bob are encountered, the outputs $A_b^i$ are uniform in the manner of \eqref{e:stipulation}:
\begin{equation}\label{e:stipulation2}
\pp(A_b^i=1|X,Y,Z,\{\text{Earlier outcomes of Alice and Charlie on their branches}\})=1/2.
\end{equation}
Conversely, when PR boxes shared with Charlie are observed, we have 
\begin{equation}\label{e:pastdet}
\pp(A_{c}^i=1|X,Y,Z,\{\text{Earlier outcomes of Alice and Charlie on their branches}\})\in \{0,1\}
\end{equation}
where the choice between 0 or 1 is determined by \eqref{e:determined}, as the conditioner will specify what Charlie's output was for the corresponding box, as well as the relevant locations on Alice's and Charlie's decision trees and thereby both inputs. Again, no-signaling considerations motivate the inclusion of Bob's setting $Y$ in the conditioner.

With the conditions we have imposed, any outputs $B_a^i$ or $B_c^i$ of the third party, Bob, will be completely determined by previously observed PR box outputs for all three parties, in the manner of \eqref{e:pastdet}. 

\subsection{Properties of the joint distribution}

We now develop some properties of the joint distribution obtained by the above method with the ordering Charlie-Alice-Bob. These properties will be used in the proofs of the constraint of the next section. They will also help demonstrate that there is only one joint distribution $\pp (\textbf{A}_b, \textbf{A}_c, \textbf{B}_a, \textbf{B}_c, \textbf{C}_a, \textbf{C}_b |X,Y,Z)$ consistent with the conditions we have set, and that it is independent of the choice of party ordering used to construct it.

\medskip

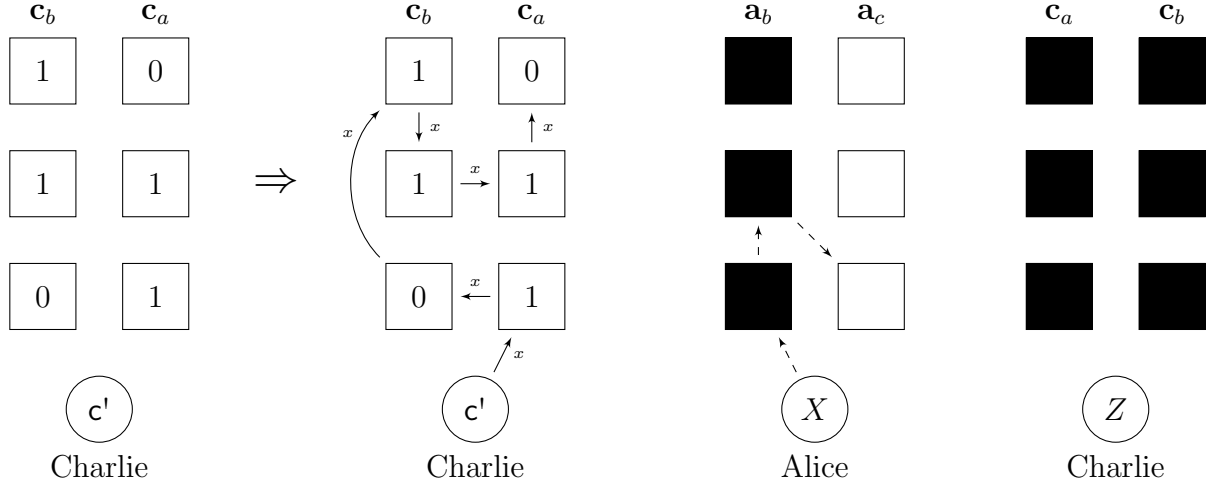
\begin{figure}[p]
    \caption{\textbf{Developing properties of the joint distribution of PR box outcomes}}
    \centering
    \begin{subfigure}[t]{0.45\textwidth}
        \centering
        \begin{tikzpicture}[>=latex']

\def\stc{1.5}
\def\cloc{2}
\def\csep{5}

        \node at (\cloc,-1+\stc/2) (cvec) {$\textbf{c}_b$};
        \node[draw,minimum size = \minsize] at (\cloc,-1) (CA3) {1};
        \node[draw,minimum size = \minsize] at (\cloc,-1-\stc) (CA2) {1};
        \node[draw,minimum size = \minsize] at (\cloc,-1-2*\stc) (CA1) {0};
        \node at (\cloc+\stc,-1+\stc/2) (cvec) {$\textbf{c}_a$};
        \node[draw,minimum size = \minsize] at (\cloc+\stc,-1) (CB3){0};
        \node[draw,minimum size = \minsize] at (\cloc+\stc,-1-\stc) (CB2){1};
        \node[draw,minimum size = \minsize] at (\cloc+\stc,-1-2*\stc) (CB1){1};
        \node[circle,draw,minimum size = \minsize] at (\cloc+0.5*\stc,-1-3*\stc) (Cset){$\mathsf{c}\textnormal{\textquotesingle}$};
        \node at (\cloc+0.5*\stc,-1-3.5*\stc) (blocka8){Charlie};
        
        \draw[shorten >=3,shorten <=3,->] (Cset) -- (CB1)node[pos = 0.5, right, text centered] {\tiny $x$};
        \draw[shorten >=3,shorten <=3,->] (CB1) -- (CA1)node[pos = 0.5, above, text centered] {\tiny $x$};
        \draw[shorten >=3,shorten <=3,->,bend left=70] (CA1) to [out=45,in=135]node[pos = 0.8, left, text centered] {\tiny $x$} (CA3);
        \draw[shorten >=3,shorten <=3,->] (CA3) -- (CA2)node[pos = 0.5, right, text centered] {\tiny $x$};
        \draw[shorten >=3,shorten <=3,->] (CA2) -- (CB2)node[pos = 0.5, above, text centered] {\tiny $x$};
        \draw[shorten >=3,shorten <=3,->] (CB2) -- (CB3)node[pos = 0.5, right, text centered] {\tiny $x$};

\node at (\cloc-\csep*.5+.4*\stc,-1-\stc) (arr){\Large$\Rightarrow$};

        \node at (\cloc-\csep,-1+\stc/2) (cvec) {$\textbf{c}_b$};
        \node[draw,minimum size = \minsize] at (\cloc-\csep,-1) (cfill1) {1};
        \node[draw,minimum size = \minsize] at (\cloc-\csep,-1-\stc) (cfill2) {1};
        \node[draw,minimum size = \minsize] at (\cloc-\csep,-1-2*\stc) (cfill3) {0};
        \node at (\cloc-\csep+\stc,-1+\stc/2) (cvec) {$\textbf{c}_a$};
        \node[draw,minimum size = \minsize] at (\cloc-\csep+\stc,-1) (cfill4){0};
        \node[draw,minimum size = \minsize] at (\cloc-\csep+\stc,-1-\stc) (cfill5){1};
        \node[draw,minimum size = \minsize] at (\cloc-\csep+\stc,-1-2*\stc) (cfill6){1};
        \node[circle,draw,minimum size = \minsize] at (\cloc-\csep+0.5*\stc,-1-3*\stc) (cfill7){$\mathsf{c}\textnormal{\textquotesingle}$};
        \node at (\cloc-\csep+0.5*\stc,-1-3.5*\stc) (b8){Charlie};

\end{tikzpicture}
\vspace{0mm}
\caption{If we randomly assign zeros and ones to all PR box outputs (left), then provided Charlie's decision tree is well formed, this will correspond to exactly one branch of Charlie's tree. This specifies the order in which the PR boxes are observed, indicated by arrows, as well as the inputs $x$ that are provided at each querying of the PR boxes (right).} \label{f:ctoc}
    \end{subfigure}
    \hfill
    \begin{subfigure}[t]{0.45\textwidth}
        \centering
\begin{tikzpicture}[>=latex']

\def\stc{1.5}
\def\cloc{2}
\def\asep{4}

        \node at (\aloc,-1+\stc/2) (abvec) {$\textbf{a}_b$};
        \node[draw,fill=black,minimum size = \minsize] at (\aloc,-1) (blocka1) {};
        \node[draw,fill=black,minimum size = \minsize] at (\aloc,-1-\stc) (blocka2) {};
        \node[draw,fill=black,,minimum size = \minsize] at (\aloc,-1-2*\stc) (blocka3) {};
        \node at (\aloc+\stc,-1+\stc/2) (abvec) {$\textbf{a}_c$};
        \node[draw,minimum size = \minsize] at (\aloc+\stc,-1) (blocka4){};
        \node[draw,minimum size = \minsize] at (\aloc+\stc,-1-\stc) (blocka5){};
        \node[draw,minimum size = \minsize] at (\aloc+\stc,-1-2*\stc) (blocka6){};
        \node[circle,draw,minimum size = \minsize] at (\aloc+0.5*\stc,-1-3*\stc) (blocka7){$X$};
        \node at (\aloc+0.5*\stc,-1-3.5*\stc) (blocka8){Alice};

        \draw[shorten >=3,shorten <=3,->,dashed] (blocka7) -- (blocka3);
                \draw[shorten >=3,shorten <=3,->,dashed] (blocka3) -- (blocka2);       
                 \draw[shorten >=3,shorten <=3,->,dashed] (blocka2) -- (blocka6);

        \node at (\aloc+\asep,-1+\stc/2) (abvec) {$\textbf{c}_a$};
        \node[draw,fill=black,minimum size = \minsize] at (\aloc+\asep,-1) (c1) {};
        \node[draw,fill=black,minimum size = \minsize] at (\aloc+\asep,-1-\stc) (c2) {$\sfrac{0}{1}$};
        \node[draw,fill=black,minimum size = \minsize] at (\aloc+\asep,-1-2*\stc) (c3) {$1$};
        \node at (\aloc+\asep+\stc,-1+\stc/2) (abvec) {$\textbf{c}_b$};        
        \node[draw,fill=black,minimum size = \minsize] at (\aloc+\asep+\stc,-1) (c4){};
        \node[draw,fill=black,minimum size = \minsize] at (\aloc+\asep+\stc,-1-\stc) (c5){};
        \node[draw,fill=black,minimum size = \minsize] at (\aloc+\asep+\stc,-1-2*\stc) (c6){};
        \node[circle,draw,minimum size = \minsize] at (\aloc+\asep+0.5*\stc,-1-3*\stc) (c7){$Z$};
        \node at (\aloc+\asep+0.5*\stc,-1-3.5*\stc) (c8){Charlie};

\end{tikzpicture}
\vspace{5mm}
        \caption{Fix a choice in $\{0,1\}^3$ for each of the three strings $\textbf{a}_b$, $\textbf{c}_a$, and $\textbf{c}_b$; these are shaded black. As Alice works through her decision tree, her path through the PR boxes (dashed arrows) is completely determined: her black boxes are already fixed, and every time she encounters a white box, only one output in $\{0,1\}$ is consistent with her input and Charlie's corresponding input and output, by \eqref{e:determined}. So Alice is confined to one branch of her tree and $\textbf{a}_c$ is fixed by the choice of $\textbf{a}_b$, $\textbf{c}_a$, and $\textbf{c}_b$.}  \label{f:ctoa}
    \end{subfigure}

    \vspace{1cm}
    \begin{subfigure}[t]{\textwidth}
    \centering
\begin{tikzpicture}[>=latex']

        \node at (\aloc,-1+\stc/2) (abvec) {$\textbf{a}_b$};
        \node[draw,fill=black,minimum size = \minsize] at (\aloc,-1) (blocka1) {};
        \node[draw,fill=black,minimum size = \minsize] at (\aloc,-1-\stc) (blocka2) {};
        \node[draw,fill=black,minimum size = \minsize] at (\aloc,-1-2*\stc) (blocka3) {};
        \node at (\aloc+\stc,-1+\stc/2) (abvec) {$\textbf{a}_c$};
        \node[draw,fill=gray!20,minimum size = \minsize] at (\aloc+\stc,-1) (blocka4){};
        \node[draw,fill=gray!20,minimum size = \minsize] at (\aloc+\stc,-1-\stc) (blocka5){};
        \node[draw,fill=gray!20,minimum size = \minsize] at (\aloc+\stc,-1-2*\stc) (blocka6){};
        \node[circle,draw,minimum size = \minsize] at (\aloc+0.5*\stc,-1-3*\stc) (blocka7){$X$};
        \node at (\aloc+0.5*\stc,-1-3.5*\stc) (blocka8){Alice};

        \node at (\aloc-\psep,-1+\stc/2) (abvec) {$\textbf{b}_c$};
        \node[draw,minimum size = \minsize] at (\aloc-\psep,-1) (b1) {};
        \node[draw,minimum size = \minsize] at (\aloc-\psep,-1-\stc) (b2) {};
        \node[draw,minimum size = \minsize] at (\aloc-\psep,-1-2*\stc) (b3) {};
        \node at (\aloc-\psep+\stc,-1+\stc/2) (abvec) {$\textbf{b}_a$};
        \node[draw,minimum size = \minsize] at (\aloc-\psep+\stc,-1) (b4){};
        \node[draw,minimum size = \minsize] at (\aloc-\psep+\stc,-1-\stc) (b5){};
        \node[draw,minimum size = \minsize] at (\aloc-\psep+\stc,-1-2*\stc) (b6){};
        \node[circle,draw,minimum size = \minsize] at (\aloc-\psep+0.5*\stc,-1-3*\stc) (b7){Y};
        \node at (\aloc-\psep+0.5*\stc,-1-3.5*\stc) (b8){Bob};

        \draw[shorten >=3,shorten <=3,->,dashed] (b7) -- (b6);
                \draw[shorten >=3,shorten <=3,->,dashed] (b6) -- (b3);       
                 \draw[shorten >=3,shorten <=3,->,dashed] (b3) -- (b4);

        \node at (\aloc+\psep,-1+\stc/2) (abvec) {$\textbf{c}_a$};
        \node[draw,fill=black,minimum size = \minsize] at (\aloc+\psep,-1) (c1) {};
        \node[draw,fill=black,minimum size = \minsize] at (\aloc+\psep,-1-\stc) (c2) {};
        \node[draw,fill=black,minimum size = \minsize] at (\aloc+\psep,-1-2*\stc) (c3) {};
        \node at (\aloc+\psep+\stc,-1+\stc/2) (abvec) {$\textbf{c}_b$};
        \node[draw,fill=black,minimum size = \minsize] at (\aloc+\psep+\stc,-1) (c4){};
        \node[draw,fill=black,minimum size = \minsize] at (\aloc+\psep+\stc,-1-\stc) (c5){};
        \node[draw,fill=black,minimum size = \minsize] at (\aloc+\psep+\stc,-1-2*\stc) (c6){};
        \node[circle,draw,minimum size = \minsize] at (\aloc+\psep+0.5*\stc,-1-3*\stc) (c7){$Z$};
        \node at (\aloc+\psep+0.5*\stc,-1-3.5*\stc) (c8){Charlie};

\end{tikzpicture}
\vspace{5mm}
        \caption{As discussed in part (b), a fixed choice for the three strings $\textbf{a}_b, \textbf{c}_a,\textbf{c}_b$ (shaded black) determines the string $\textbf{a}_c$ (shaded grey). Hence the branches of Alice's and Charlie's decision trees are uniquely determined, as are all of their PR box inputs and outputs. Then Bob's decision tree will dictates a unique path through his PR box outputs with his outcomes determined by \eqref{e:determined}, fixing $\textbf{b}_a$ and $\textbf{b}_c$.} \label{f:catob}
    \end{subfigure}
\end{figure}

\noindent 1. {\it The marginal distribution of all of Charlie's PR box outcomes is uniform}. That is,
\begin{equation}\label{e:charlieunif}
\pp(\textbf{c}_a, \textbf{c}_b|X,Y,Z) = \frac{1}{2^{2n}} \quad \text{for all } \textbf{c}_a, \textbf{c}_b \in \{0,1\}^n,
\end{equation}
where Charlie shares $n$ PR boxes with Alice and $n$ PR boxes with Bob. Above, we are using lower case letters to represent values that can be taken by random variables represented by the corresponding upper case letters, as well as the shorthand $\pp(\textbf{c}_a)$ for $\pp(\textbf{C}_a=\textbf{c}_a)$. We can see that \eqref{e:charlieunif} follows from \eqref{e:stipulation} so long as Charlie's decision tree is well-formed, in the sense that branches never direct you to query a $C_q^i$ location that has been visited earlier on the branch, and always give you an instruction for where to go if you get 0 or 1: if we fix a possibility $(\textbf{C}_a, \textbf{C}_b)=(\textbf{c}_a,\textbf{c}_b)\in \{0,1\}^{2n}$ and assign the corresponding 0 or 1 value to all $C_q^i$ locations (see Fig.~\ref{f:ctoc}), then given the value of $Z$, one can work through the decision tree corresponding to this assignment and end up at the end of exactly one branch. The condition \eqref{e:stipulation} then ensures that the probability that this branch occurs is $2n$ iterates of $1/2$ multiplied together.

\medskip

\noindent 2. \textit{The joint distribution of $\textbf{A}_b$, $\textbf{C}_a$, and $\textbf{C}_b$ is uniform}:
\begin{equation}\label{e:threeindep}
\pp(\textbf{a}_b, \textbf{c}_a, \textbf{c}_b|X,Y,Z) = \frac{1}{2^{3n}} \quad \text{for all } \textbf{a}_b, \textbf{c}_a, \textbf{c}_b \in \{0,1\}^n.
\end{equation}
Note that equation \eqref{e:threeindep} allows us to write $\pp(\textbf{a}_b|X,Y,Z)=\sum_{\textbf{c}_a, \textbf{c}_b}\pp(\textbf{a}_b, \textbf{c}_a, \textbf{c}_b|X,Y,Z) = 2^{-n}$, which together with \eqref{e:charlieunif} implies 
\begin{equation}\label{e:threeindep1}
\pp({\bf a}_b, {\bf c}_a, {\bf c}_b|X,Y,Z) = \pp({\bf a}_b|X,Y,Z) \pp({\bf c}_a, {\bf c}_b|X,Y,Z) \quad \text{for all } \textbf{a}_b, \textbf{c}_a, \textbf{c}_b \in \{0,1\}^n
\end{equation}
where $\pp(\textbf{a}_b|X,Y,Z) = 2^{-n}$ and $\pp(\textbf{c}_a, \textbf{c}_b|X,Y,Z)=2^{-2n}$ for all choices of $\textbf{a}_b,\textbf{c}_a,\textbf{c}_b$. 

To see why \eqref{e:threeindep} holds, fix any $\textbf{a}_b\in \{0,1\}^n$ and $(\textbf{c}_a,\textbf{c}_b)\in \{0,1\}^{2n}$. The choice of $(\textbf{c}_a,\textbf{c}_b)$ corresponds to a unique branch of Charlie's decision tree as discussed in the previous point. Then we work through Alice's decision tree (see Fig.~\ref{f:ctoa}), with choices of $A_b^i$ set by the $i$th value of $\textbf{a}_b$ and choices of $A_c^i$ fixed through \eqref{e:pastdet} by Charlie's output (the $i$th entry of $\textbf{c}_b$), Alice's input (given by her location in her decision tree), and Charlie's input (given by the location of $C_a^i$ in the unique branch of Charlie's tree corresponding to $(\textbf{c}_a,\textbf{c}_b)$). If Alice's tree is well formed, this all uniquely determines a single branch of her tree. The probability of Alice observing this branch, conditioned on Charlie observing $(\textbf{C}_a, \textbf{C}_b )= (\textbf{c}_a,\textbf{c}_b)$, can be computed by multiplying $n$ copies of $1/2$ for the $A_b^i$ branch points (by \eqref{e:stipulation2}) and $n$ copies of 1 for the $A_c^i$ branch points (by \eqref{e:pastdet}). Since by \eqref{e:charlieunif} the probability that Charlie observes $(\textbf{C}_a, \textbf{C}_b )= (\textbf{c}_a,\textbf{c}_b)$ is $2^{-2n}$, the final probability is then $2^{-3n}$, yielding \eqref{e:threeindep}.

The above argument implies a useful result:
\begin{eqnarray}
\pp(\textbf{a}_b,\textbf{a}_c,\textbf{c}_a,\textbf{c}_b|X,Y,Z)&=&\pp({\bf a}_b|X,Y,Z) \pp({\bf c}_a, {\bf c}_b|X,Y,Z)\llbracket \textbf{a}_c=\textbf{A}_c({\bf a}_b,{\bf c}_a, {\bf c}_b, X,Z)\rrbracket\notag\\
&=&\frac{1}{2^n} \frac{1}{2^{2n}}\llbracket \textbf{a}_c=\textbf{A}_c({\bf a}_b,{\bf c}_a, {\bf c}_b, X,Z)\rrbracket\label{e:parametrizing}
\end{eqnarray}
Above we introduce two notations: First, the expression $\llbracket \cdots \rrbracket$ represents the function that evaluates to 1 if the contained statement is true and 0 if the contained statement is false. Second, we define a new function $\textbf{A}_c(\cdot,\cdot,\cdot,\cdot,\cdot)$ as follows. As described in the previous paragraph and Fig.~\ref{f:ctoa}, when given a fixed value of $\textbf{a}_b$, $\textbf{c}_a$, $\textbf{c}_b$, and a choice of settings $X$ and $Z$, there is only one possible value for $\bf{a}_c$ consistent with Alice's and Charlie's decision trees. $\textbf{A}_c(\cdot,\cdot,\cdot,\cdot,\cdot)$ is the function that returns this value when given the inputs $\textbf{a}_b$, $\textbf{c}_a$, $\textbf{c}_b,x, z$.

Equation \eqref{e:threeindep1} is an interesting feature of the joint distribution of PR box outputs -- no matter the strategy, ${\bf A}_b$ is free even given Charlie's outputs. This result and \eqref{e:parametrizing} are closely related to the ``parametrizing the randomness" arguments of \cite{chao:2017}, and will be used in the upcoming proofs. 

\medskip

\noindent 3. \textit{Given fixed values of ${\bf A}_b={\bf a}_b$, ${\bf C}_a={\bf c}_a$, and ${\bf C}_b={\bf c}_b$, Bob's two strings $\textbf{B}_a$ and $\textbf{B}_c$ are completely determined}:
\begin{equation}
\pp(\textbf{b}_a,\textbf{b}_c|\textbf{a}_b, \textbf{c}_a, \textbf{c}_b,X,Y,Z) \in \{0,1\}.
\end{equation}
To see why this is so, recall that the choices ${\bf A}_b={\bf a}_b$, ${\bf C}_a={\bf c}_a$, and ${\bf C}_b={\bf c}_b$ fix $\textbf{A}_c$ uniquely and determine both Charlie's and Alice's branches of their respective decision trees. Then we can see that as Bob works through his decision tree (see Fig.~\ref{f:catob}), his output to every PR box is fixed by a condition similar to that of \eqref{e:pastdet} because the other parties' outputs and inputs are set. This leaves Bob with one possible branch of his tree corresponding to a single value of $(\textbf{b}_a,\textbf{b}_c)$.

\subsection{Consistency of the construction of the joint distribution}\label{s:consist}

It is now clear that there is a unique joint distribution $\pp (\textbf{A}_b, \textbf{A}_c, \textbf{B}_a, \textbf{B}_c, \textbf{C}_a, \textbf{C}_b |X,Y,Z)$ consistent with our stipulations: a collection of strings $\textbf{a}_b, \textbf{a}_c, \textbf{b}_a, \textbf{b}_c, \textbf{c}_a, \textbf{c}_b$ has probability $2^{-3n}$ if the strings $\textbf{a}_c,\textbf{b}_a,\textbf{b}_c$ are precisely those that are determined by the choice $\textbf{a}_b,\textbf{c}_a,\textbf{c}_b$, and zero otherwise. This distribution is derived based on an arbitrary ordering of the parties -- Charlie, then Alice, then Bob. We now show that we obtain the same distribution if we choose any different ordering of parties. It suffices to show the same distribution is obtained if we either 1) swap the order of the first and second parties (Charlie and Alice), or 2) swap the order of the second and third parties (Alice and Bob), as iterations of these two swapping operations can induce any ordering of parties. 

\begin{figure}\caption{\textbf{Switching the order of parties leads to the same joint distribution.}}
\centering
    \begin{subfigure}[t]{1\textwidth}
    \centering 
\begin{tikzpicture}
\def\stc{1.5}
\def\aloc{-5}
\def\asep{4}
\def\plsep{10}
\def\minsize{25}

        \node at (\aloc,-1+\stc/2) (abvec) {$\textbf{a}_b$};
        \node[draw,fill=black,minimum size = \minsize] at (\aloc,-1) (blocka1) {};
        \node[draw,fill=black,minimum size = \minsize] at (\aloc,-1-\stc) (blocka2) {};
        \node[draw,fill=black,,minimum size = \minsize] at (\aloc,-1-2*\stc) (blocka3) {};
        \node at (\aloc+\stc,-1+\stc/2) (abvec) {$\textbf{a}_c$};
        \node[draw,minimum size = \minsize] at (\aloc+\stc,-1) (blocka4){};
        \node[draw,minimum size = \minsize] at (\aloc+\stc,-1-\stc) (blocka5){};
        \node[draw,minimum size = \minsize] at (\aloc+\stc,-1-2*\stc) (blocka6){};
        \node[circle,draw,minimum size = \minsize] at (\aloc+0.5*\stc,-1-3*\stc) (blocka7){$X$};
        \node at (\aloc+0.5*\stc,-1-3.5*\stc) (blocka8){Alice};
        
                \draw[shorten >=3,shorten <=3,->,dashed] (blocka7) -- (blocka3);
                \draw[shorten >=3,shorten <=3,->,dashed] (blocka3) -- (blocka6);
                \draw[shorten >=3,shorten <=3,->,dashed] (blocka6) -- (blocka5);
                \draw[shorten >=3,shorten <=3,->,dashed] (blocka5) -- (blocka4);
                \draw[shorten >=3,shorten <=3,->,dashed] (blocka4) -- (blocka2);
                \draw[shorten >=3,shorten <=3,->,dashed] (blocka2) -- (blocka1);

        \node at (\aloc+\asep,-1+\stc/2) (abvec) {$\textbf{c}_a$};
        \node[draw,fill=black,minimum size = \minsize] at (\aloc+\asep,-1) (c1) {};
        \node[draw,fill=black,minimum size = \minsize] at (\aloc+\asep,-1-\stc) (c2) {};
        \node[draw,fill=black,minimum size = \minsize] at (\aloc+\asep,-1-2*\stc) (c3) {};
        \node at (\aloc+\asep+\stc,-1+\stc/2) (abvec) {$\textbf{c}_b$};        
        \node[draw,fill=black,minimum size = \minsize] at (\aloc+\asep+\stc,-1) (c4){};
        \node[draw,fill=black,minimum size = \minsize] at (\aloc+\asep+\stc,-1-\stc) (c5){};
        \node[draw,fill=black,minimum size = \minsize] at (\aloc+\asep+\stc,-1-2*\stc) (c6){};
        \node[circle,draw,minimum size = \minsize] at (\aloc+\asep+0.5*\stc,-1-3*\stc) (c7){$Z$};
        \node at (\aloc+\asep+0.5*\stc,-1-3.5*\stc) (c8){Charlie};

                \draw[shorten >=3,shorten <=3,->] (c7) -- (c6);
                \draw[shorten >=3,shorten <=3,->] (c6) -- (c5);       
                 \draw[shorten >=3,shorten <=3,->] (c5) -- (c3);       
                 \draw[shorten >=3,shorten <=3,->] (c3) -- (c2); 
                 \draw[shorten >=3,shorten <=3,->] (c2) -- (c1);
                 \draw[shorten >=3,shorten <=3,->] (c1) -- (c4);
        
\node at (\aloc+.5*\psep+.3*\stc+\asep,-1-\stc) (argh){\huge$\Leftrightarrow$};

                \node at (\aloc+\plsep,-1+\stc/2) (abvec2) {$\textbf{a}_b$};
        \node[draw,fill=black,minimum size = \minsize] at (\aloc+\plsep,-1) (a1) {};
        \node[draw,fill=black,minimum size = \minsize] at (\aloc+\plsep,-1-\stc) (a2) {};
        \node[draw,fill=black,,minimum size = \minsize] at (\aloc+\plsep,-1-2*\stc) (a3) {};
        \node at (\aloc+\stc+\plsep,-1+\stc/2) (abvec2) {$\textbf{a}_c$};
        \node[draw,fill=black,minimum size = \minsize] at (\aloc+\stc+\plsep,-1) (a4){};
        \node[draw,fill=black,minimum size = \minsize] at (\aloc+\stc+\plsep,-1-\stc) (a5){};
        \node[draw,fill=black,minimum size = \minsize] at (\aloc+\stc+\plsep,-1-2*\stc) (a6){};
        \node[circle,draw,minimum size = \minsize] at (\aloc+0.5*\stc+\plsep,-1-3*\stc) (a7){$X$};
        \node at (\aloc+0.5*\stc+\plsep,-1-3.5*\stc) (a8){Alice};
        
                \draw[shorten >=3,shorten <=3,->] (a7) -- (a3);
                \draw[shorten >=3,shorten <=3,->] (a3) -- (a6);
                \draw[shorten >=3,shorten <=3,->] (a6) -- (a5);
                \draw[shorten >=3,shorten <=3,->] (a5) -- (a4);
                \draw[shorten >=3,shorten <=3,->] (a4) -- (a2);
                \draw[shorten >=3,shorten <=3,->] (a2) -- (a1);

        \node at (\aloc+\asep+\plsep,-1+\stc/2) (abvec) {$\textbf{c}_a$};
        \node[draw,minimum size = \minsize] at (\aloc+\asep+\plsep,-1) (c1) {};
        \node[draw,minimum size = \minsize] at (\aloc+\asep+\plsep,-1-\stc) (c2) {};
        \node[draw,minimum size = \minsize] at (\aloc+\asep+\plsep,-1-2*\stc) (c3) {};
        \node at (\aloc+\asep+\plsep+\stc,-1+\stc/2) (abvec) {$\textbf{c}_b$};        
        \node[draw,fill=black,minimum size = \minsize] at (\aloc+\asep+\stc+\plsep,-1) (c4){};
        \node[draw,fill=black,minimum size = \minsize] at (\aloc+\asep+\stc+\plsep,-1-\stc) (c5){};
        \node[draw,fill=black,minimum size = \minsize] at (\aloc+\asep+\stc+\plsep,-1-2*\stc) (c6){};
        \node[circle,draw,minimum size = \minsize] at (\aloc+\asep+0.5*\stc+\plsep,-1-3*\stc) (c7){$Z$};
        \node at (\aloc+\asep+0.5*\stc+\plsep,-1-3.5*\stc) (c8){Charlie};
        
                \draw[shorten >=3,shorten <=3,->,dashed] (c7) -- (c6);
                \draw[shorten >=3,shorten <=3,->,dashed] (c6) -- (c5);       
                 \draw[shorten >=3,shorten <=3,->,dashed] (c5) -- (c3);       
                 \draw[shorten >=3,shorten <=3,->,dashed] (c3) -- (c2); 
                 \draw[shorten >=3,shorten <=3,->,dashed] (c2) -- (c1);
                 \draw[shorten >=3,shorten <=3,->,dashed] (c1) -- (c4);
       
\end{tikzpicture}
\caption{If the party ordering for constructing the distribution is Alice-Charlie-Bob (right), the first step is to fix an arbitrary choice for $\textbf{a}_b,\textbf{a}_c,\textbf{c}_b$, denoted by black shading. The choice of $\textbf{a}_b$ and $\textbf{a}_c$ completely determines Alice's path on her decision tree, which is represented by solid arrows. One then works through Charlie's decision tree (dashed arrows), filling in the $\textbf{c}_a$ entries along the way. Now suppose you take this collection of all four strings $\textbf{a}_b,\textbf{a}_c,\textbf{c}_a,\textbf{c}_b$ and erase $\textbf{a}_c$ (left). First, the strings $\textbf{c}_a,\textbf{c}_b$ must induce the same path on Charlie's decision tree as before,  now represented by solid arrows, since there is only one branch of a well-formed decision tree consistent with a fixed collection of PR box outputs (recall Fig.~\ref{f:ctoc}). The next step in the Charlie-Alice-Bob ordering is to fill in $\textbf{A}_c$ by working through Alice's decision tree. With some thought, we see that we must remain on the same branch of the tree and recover the original $\textbf{a}_c$: as Alice encounters $A_c^i$ locations, both Alice's and Charlie's PR box inputs will be the same as they were at the corresponding location of the above-right diagram, and so the choice between the relationship $A_c^i=C_a^i$ or $A_c^i\ne C_a^i$ governed by \eqref{e:determined} will hold as before. This one-to-one correspondence of the possibilities for filling $\textbf{a}_c$ as the fourth string, versus filling $\textbf{c}_a$ as the fourth string, implies that a choice $\textbf{a}_b,\textbf{a}_c,\textbf{c}_a,\textbf{c}_b$ is consistent with an Alice-Charlie-Bob party ordering for constructing the joint distribution if and only if it is consistent with a Charlie-Alice-Bob party ordering for constructing the joint distribution. 
}\label{f:acswitch}
\end{subfigure}
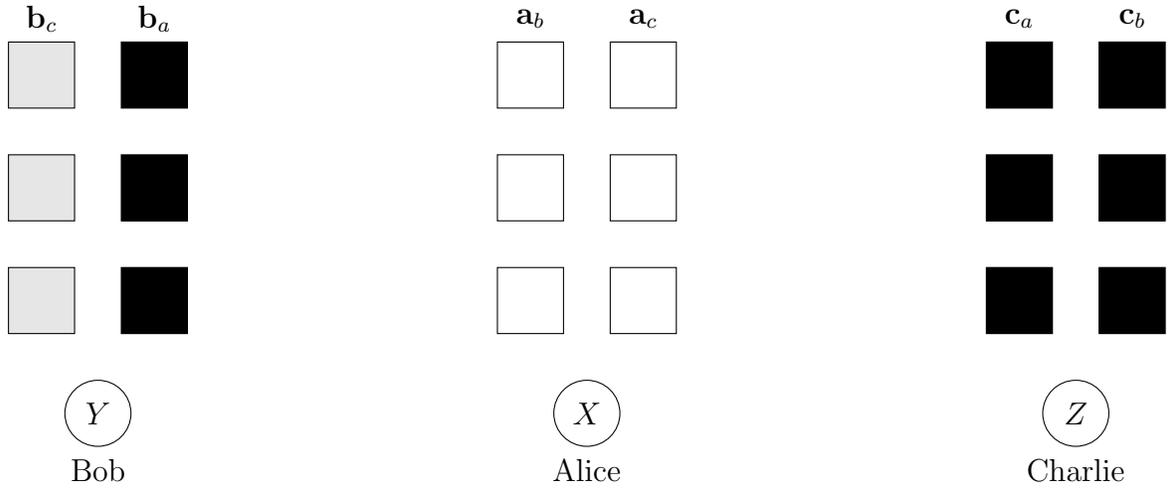
    \begin{subfigure}[t]{\textwidth}
    \centering
\vspace{2mm}
\begin{tikzpicture}[>=latex']

        \node at (\aloc,-1+\stc/2) (abvec) {$\textbf{a}_b$};
        \node[draw,minimum size = \minsize] at (\aloc,-1) (blocka1) {};
        \node[draw,minimum size = \minsize] at (\aloc,-1-\stc) (blocka2) {};
        \node[draw,minimum size = \minsize] at (\aloc,-1-2*\stc) (blocka3) {};
        \node at (\aloc+\stc,-1+\stc/2) (abvec) {$\textbf{a}_c$};
        \node[draw,minimum size = \minsize] at (\aloc+\stc,-1) (blocka4){};
        \node[draw,minimum size = \minsize] at (\aloc+\stc,-1-\stc) (blocka5){};
        \node[draw,minimum size = \minsize] at (\aloc+\stc,-1-2*\stc) (blocka6){};
        \node[circle,draw,minimum size = \minsize] at (\aloc+0.5*\stc,-1-3*\stc) (blocka7){$X$};
        \node at (\aloc+0.5*\stc,-1-3.5*\stc) (blocka8){Alice};

        \node at (\aloc-\psep,-1+\stc/2) (abvec) {$\textbf{b}_c$};
        \node[draw,fill=gray!20,minimum size = \minsize] at (\aloc-\psep,-1) (b1) {};
        \node[draw,fill=gray!20,minimum size = \minsize] at (\aloc-\psep,-1-\stc) (b2) {};
        \node[draw,fill=gray!20,minimum size = \minsize] at (\aloc-\psep,-1-2*\stc) (b3) {};
        \node at (\aloc-\psep+\stc,-1+\stc/2) (abvec) {$\textbf{b}_a$};
        \node[draw,fill=black,minimum size = \minsize] at (\aloc-\psep+\stc,-1) (b4){};
        \node[draw,fill=black,minimum size = \minsize] at (\aloc-\psep+\stc,-1-\stc) (b5){};
        \node[draw,fill=black,minimum size = \minsize] at (\aloc-\psep+\stc,-1-2*\stc) (b6){};
        \node[circle,draw,minimum size = \minsize] at (\aloc-\psep+0.5*\stc,-1-3*\stc) (b7){$Y$};
        \node at (\aloc-\psep+0.5*\stc,-1-3.5*\stc) (b8){Bob};

        \node at (\aloc+\psep,-1+\stc/2) (abvec) {$\textbf{c}_a$};
        \node[draw,fill=black,minimum size = \minsize] at (\aloc+\psep,-1) (c1) {};
        \node[draw,fill=black,minimum size = \minsize] at (\aloc+\psep,-1-\stc) (c2) {};
        \node[draw,fill=black,minimum size = \minsize] at (\aloc+\psep,-1-2*\stc) (c3) {};
        \node at (\aloc+\psep+\stc,-1+\stc/2) (abvec) {$\textbf{c}_b$};
        \node[draw,fill=black,minimum size = \minsize] at (\aloc+\psep+\stc,-1) (c4){};
        \node[draw,fill=black,minimum size = \minsize] at (\aloc+\psep+\stc,-1-\stc) (c5){};
        \node[draw,fill=black,minimum size = \minsize] at (\aloc+\psep+\stc,-1-2*\stc) (c6){};
        \node[circle,draw,minimum size = \minsize] at (\aloc+\psep+0.5*\stc,-1-3*\stc) (c7){$Z$};
        \node at (\aloc+\psep+0.5*\stc,-1-3.5*\stc) (c8){Charlie};

\end{tikzpicture}
\caption{If the order is Charlie-Bob-Alice, then $\textbf{b}_a,\textbf{c}_a,\textbf{c}_b$ (black) are chosen from a uniform distribution. Once these three strings are fixed, then $\textbf{b}_c$ is determined first (gray), followed by Alice's strings $\textbf{a}_b$ and $\textbf{a}_c$ (white). Suppose now we take the so-obtained string $\textbf{a}_b$ with the original choices of $\textbf{c}_a$ and $\textbf{c}_b$ as a starting point for determining the remaining strings according to the Charlie-Alice-Bob order. Then first $\textbf{a}_c$ will be recovered according to the scheme in Fig.~\ref{f:catob}, followed by the Bob strings. With some thought one sees this will result in the same six strings as originally determined in the Charlie-Bob-Alice ordering.}\label{f:abswitch}
        \end{subfigure}
        
\end{figure}
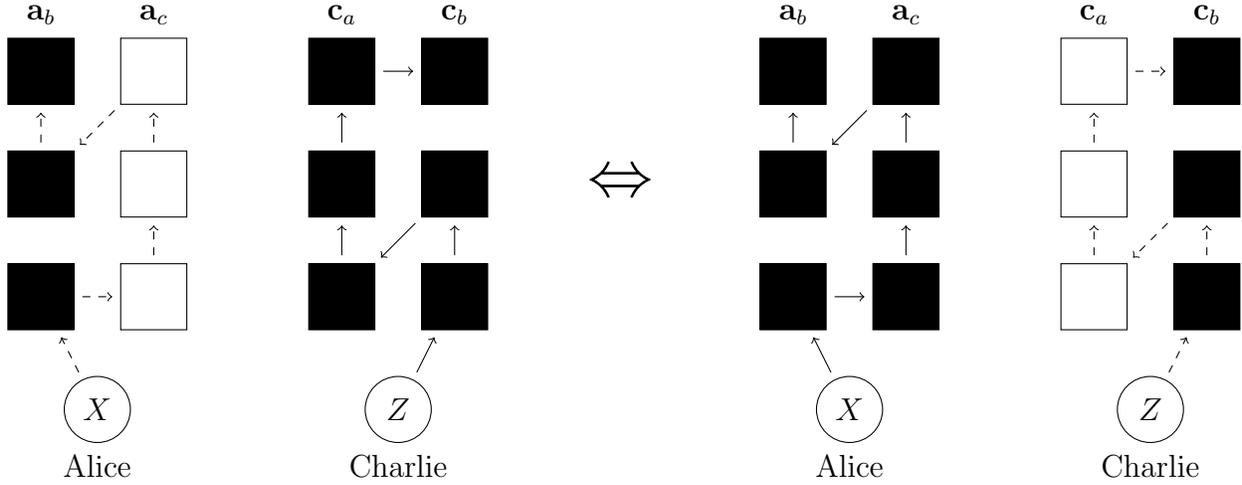

First consider switching the order of Charlie and Alice. The equivalent condition to \eqref{e:threeindep} is then that $\textbf{A}_b,\textbf{A}_c,\textbf{C}_b$ are uniformly distributed, and then the remaining three strings $\textbf{C}_a, \textbf{B}_a,\textbf{B}_c$ can be obtained as a function of the first three. Starting with an arbitrary choice of $\textbf{a}_b,\textbf{a}_c,\textbf{c}_b$, the procedure that determines the remaining strings is to first calculate $\textbf{c}_a$ by working through Charlie's decision tree. As illustrated in Figure \ref{f:acswitch}, this process yields a $\textbf{c}_a$ with a particular property: had we randomly selected this $\textbf{c}_a$ and combined it with our earlier choice of $\textbf{a}_b$ and $\textbf{c}_b$ as a starting point for filling the PR boxes with the original Charlie-Alice-Bob ordering, then in determining $\textbf{a}_c$ in the manner of Fig.~\ref{f:ctoa}, we would end up with the same choice for $\textbf{a}_c$ that we started with earlier. This implies that the same $2^{3n}$ joint possibilities for $\textbf{C}_a, \textbf{C}_b, \textbf{A}_b,\textbf{A}_c$ are obtained whether we start with an arbitrary choice of $\textbf{a}_b,\textbf{a}_c,\textbf{c}_b$ that then determines a unique $\textbf{c}_a$ (Alice-Charlie-Bob ordering), or start with an arbitrary choice of $\textbf{a}_b,\textbf{c}_a,\textbf{c}_b$ that then determines a unique $\textbf{a}_c$ (Charlie-Alice-Bob ordering). Bob's strings are then determined the same way in either ordering of parties in the manner of Figure \ref{f:catob}, and so the same distribution $\pp (\textbf{A}_b, \textbf{A}_c, \textbf{B}_a, \textbf{B}_c, \textbf{C}_a, \textbf{C}_b |X,Y,Z)$ is obtained. 

Now, consider instead switching the order of Alice and Bob, so the new order is Charlie-Bob-Alice. As shown in Figure \ref{f:abswitch}, this entails a situation where $\textbf{C}_a,\textbf{C}_b,\textbf{B}_a$ are uniformly distributed with other strings obtained as functions of these. While the situation is a little more complicated now, the same general idea holds: a fixed value for $\textbf{c}_a,\textbf{c}_b,\textbf{b}_a$ determines a choice for $\textbf{a}_b$ (along with the other two strings), but if we start instead with this choice of $\textbf{a}_b$ along with $\textbf{c}_a,\textbf{c}_b$ and determine the remaining three strings according to the Charlie-Alice-Bob order, we will get the same six strings either way. This correspondence implies the same distribution $\pp (\textbf{A}_b, \textbf{A}_c, \textbf{B}_a, \textbf{B}_c, \textbf{C}_a, \textbf{C}_b |X,Y,Z)$ is obtained.


\subsection{No-signaling properties of the observed outcomes}\label{s:nosigs}

We now demonstrate no-signaling-type properties of the distribution $P(A,B,C|X,Y,Z)$ of observed outcomes. We can freely apply the results \eqref{e:charlieunif}-\eqref{e:parametrizing} to different subsets of parties without fearing inconsistency by invoking the independence of party ordering in the construction of the joint distribution demonstrated in Subsection \ref{s:consist}. 

First we confirm that a single party's output is independent of the other two parties' settings. Let $A(\cdot,\cdot,\cdot)$ represent the function Alice applies to $\mathbf{A}_b,\mathbf{A}_c,X$ to obtain her final observed outcome $A \in \{0,\text{+}\}$. Then we can write
\begin{equation}\label{e:nosigopset}
\pp(A=\text{+}|xyz) = \sum_{\mathbf{a}_b,\mathbf{a}_c}\pp (\mathbf{a}_b, \mathbf{a}_c|xyz)\llbracket A(\textbf{a}_b,\textbf{a}_c,x)=\text{+}\rrbracket.
\end{equation}
Applying \eqref{e:charlieunif} to Alice's marginal distribution, $\pp (\mathbf{a}_b, \mathbf{a}_c|xyz)=2^{-2n}$ for all choices of $\mathbf{a}_b, \mathbf{a}_c$ regardless of $y$ and $z$, and since the predicate in brackets $\llbracket \cdots \rrbracket$ also does not depend on $y$ or $z$, the value of the sum in \eqref{e:nosigopset}  -- and thus $\pp(A=\text{+}|xyz)$ -- is independent of these remote settings. Note that while the form of the function $A(\cdot,\cdot,\cdot)$ depends on Alice's particular strategy, it is independent of the other parties' decision trees and final output functions. Hence we can say that not only is Alice's outcome probability independent of Bob's and Charlie's settings, but it is also independent of their strategies. This result naturally holds for any other choice of parties.

We also can confirm that the joint outcome distribution of two parties is independent of the settings and strategy of the third, which is another flavor of no-signaling. Suppose we have predicate $Q$ involving only two parties -- for instance, $Q$ could be ``Alice and Charlie's final outputs $A$ and $C$ are equal." Whether $Q$ occurs on a given setting $x$ and $z$ is a function of Alice and Charlie's PR box outputs, and so
\begin{equation}\label{e:nosig}
\pp(Q|xyz) = \sum_{\textbf{a}_b,\textbf{a}_c,\textbf{c}_a,\textbf{c}_b:\text{ $Q$ is true}} \pp(\textbf{a}_b,\textbf{a}_c,\textbf{c}_a,\textbf{c}_b|xyz).
\end{equation}
Applying \eqref{e:parametrizing} to $\pp(\textbf{a}_b,\textbf{a}_c,\textbf{c}_a,\textbf{c}_b|xyz)$ makes clear the independence from Bob's setting and strategy. 


\medskip

\medskip

\section{Deriving an Inequality for PR-box-Simulable Distributions}\label{s:CR}

In our (3,2,2) Bell scenario, if the observed random variables $A$, $B$, and $C$ are functions of an underlying system of PR boxes as explored in the previous sections, it is possible to derive a Bell-like inequality constraint on the probability distribution $\pp(A,B,C|X,Y,Z)$. Inspired by the game of Chao and Reichardt \cite{chao:2017}, we analyze a closely related experimental scenario in which the settings $X$, $Y$, and $Z$ are equiprobable and a score is assigned to each possible outcome of the experiment. Table \ref{t:chaogame} displays the scores for each of the $64$ setting/outcome possibilities.

\begin{table}[h!]\centering\caption{\textbf{A Bell function for the $(3,2,2)$ scenario.} Blank cells correspond to the value 0.}\label{t:chaogame}
\begin{tabular}{|cc|cccccccc|c|}
\hline
&& \multicolumn{8}{c|}{Outcomes $ABC$}&Nonzero\\
 & & +++&++0&+0+&+00&0++&0+0&00+&000&Condition\\
\hline
&$\mathsf{abc}$&&2&&2&2&&2&&$A\ne C$\\
&$\mathsf{abc\textnormal{\textquotesingle}}$&&&1&1&1&1&&&$A\ne B$\\
&$\mathsf{ab\textnormal{\textquotesingle}c}$&&2&&2&2&&2&&$A\ne C$\\
Setting&$\mathsf{ab\textnormal{\textquotesingle}c\textnormal{\textquotesingle}}$&&&1&1&1&1&&&$A \ne B$\\
$XYZ$&$\mathsf{a\textnormal{\textquotesingle}bc}$&&&&&&&&&\\
&$\mathsf{a\textnormal{\textquotesingle}bc\textnormal{\textquotesingle}}$&&1&1&&1&&&1&even num. of ``+"\\
&$\mathsf{a\textnormal{\textquotesingle}b\textnormal{\textquotesingle}c}$&&&&&&&&&\\
&$\mathsf{a\textnormal{\textquotesingle}b\textnormal{\textquotesingle}c\textnormal{\textquotesingle}}$&1&&&1&&1&1&&odd num. of ``+"\\
\hline
\end{tabular}
\end{table}

This assignment of scores to outcomes -- a \textit{Bell function} -- effectively replaces the biased settings probabilities of Chao and Reichardt with a heavier weighting of 2 on some of the outcomes. Let $F$ be the random variable that outputs the observed value of the Bell function, so $F$ is a function of $A,B,C$ and $X,Y,Z$. For an experiment with equiprobable settings, we can write the expected value of $F$ as:
\begin{eqnarray}\label{e:expoff}
&& \sum_{abc,xyz}\pp(abc|xyz)\pp(xyz)F(abc,xyz)\notag\\
&=& \frac{1}{8}\sum_{abc,xyz}\pp(abc|xyz)F(abc,xyz)\notag\\
&=&\frac{1}{8}\big[\pp(A\ne C|\mathsf{abc})\times2 +\pp(A\ne B|\mathsf{abc}\textnormal{\textquotesingle})\times1+\pp(A\ne C|\mathsf{ab\textnormal{\textquotesingle}c})\times2+\pp(A\ne B|\mathsf{ab\textnormal{\textquotesingle}c\textnormal{\textquotesingle}})\times1\notag\\
&&+\pp(A\oplus B\oplus C=0|\mathsf{a\textnormal{\textquotesingle}bc\textnormal{\textquotesingle}})\times1+\pp(A\oplus B\oplus C=1|\mathsf{a\textnormal{\textquotesingle}b\textnormal{\textquotesingle}c\textnormal{\textquotesingle}})\times1\big].\label{e:uncompact}
\end{eqnarray}
For nonsignaling distributions, \eqref{e:uncompact} can be written in the compact form
\begin{equation}\label{e:compact}
E(F)=\frac{1}{8}\left[4\times \pp(A\ne C|\mathsf{ac})+\sum_{\mathsf{y}\in\{\mathsf{b},\mathsf{b}\textnormal{\textquotesingle}\}}\big[\pp(A\ne B|\mathsf{ay})+\pp(A\oplus B \oplus C = \mathsf{y}|\mathsf{a}\textnormal{\textquotesingle}\mathsf{y}\mathsf{c}\textnormal{\textquotesingle})\big]\right].
\end{equation}
If the probability distribution $\pp(A,B,C|X,Y,Z)$ is induced by a system of PR boxes, the following inequality holds for the expected value of $F$:
\begin{equation}\label{e:ourineq}
E(F) \ge \frac{1}{8}
\end{equation}
Note this inequality is tight: the trivial strategy that outputs $A=B=C=\text{+}$ for all settings independently of PR box outputs -- which is just a local deterministic strategy -- achieves the bound. The inequality can be violated by a quantum-achievable distribution described in \cite{chao:2017}. We spend the rest of this section showing that \eqref{e:ourineq} holds for distributions induced by underlying networks of PR boxes. 

The method of proof, following the contours of the argument of Chao and Reichardt, is to show that any overall strategy $S$ -- consisting of decision trees and outcome functions for each of Alice, Bob, and Charlie -- can be modified into a simpler overall strategy $S'$ which differs only in Alice's behavior on setting $X=\mathsf{a}$, for which the settings-conditional probabilities of getting $F=0$ scores in Table \ref{t:chaogame} are not too far changed from those of $S$. Then $S'$ is further modified into a strategy $S''$ that has a deterministic output for $A$ when the setting $X=\mathsf{a}$, where again the probabilities of $F=0$ are not too far changed. Then it is pointed out that a strategy like $S''$ with a deterministic output on a given setting is constrained in its ability to win a nonlocal game, so the probability of $F=0$ for $S''$ can be bounded below, and then related to a lower bound on the expectation of $F$ for strategy $S$.

\subsection{Modifying a general strategy $S$ to a simpler strategy $S'$}

For the first part of the argument, we show how\, given a strategy $S$, to obtain a new strategy $S'$ that performs better in obtaining a smaller value, or at least an equal value, with the first and third terms in brackets in \eqref{e:uncompact}. The new strategy $S'$ will also have the property that Alice's output $A$ is independent of the outputs of PR boxes shared with Bob.

To construct $S'$, consider the setting configuration where $XYZ=\mathsf{abc}$ (the argument is the same for $Y=\mathsf{b}\textnormal{\textquotesingle}$), and let us use the notation $A(\cdot,\cdot,\cdot)$ and $C(\cdot,\cdot,\cdot)$ to represent the functions that Alice and Charlie use to determine their final outputs $A$ and $C$ from PR box outputs and the settings $X$ and $Z$. Then we can use \eqref{e:parametrizing} to write 
\begin{eqnarray}
&&\pp_S(A=C|\mathsf{abc})\notag\\
&=& \sum_{\textbf{a}_b,\textbf{a}_c,\textbf{c}_a,\textbf{c}_b} \pp_S(\textbf{a}_b,\textbf{a}_c,\textbf{c}_a,\textbf{c}_b|\mathsf{abc})\llbracket A(\textbf{a}_b,\textbf{a}_c,\mathsf{a}) = C(\textbf{c}_a,\textbf{c}_b,\mathsf{c})\rrbracket\notag\\
&=& \sum_{\textbf{a}_b,\textbf{a}_c,\textbf{c}_a,\textbf{c}_b}  \pp_S(\textbf{a}_b|\mathsf{abc})\pp_S(\textbf{c}_a,\textbf{c}_b|\mathsf{abc})\llbracket \textbf{a}_c = \textbf{A}_c^{S}(\textbf{a}_b,\textbf{c}_a,\textbf{c}_b,\mathsf{a},\mathsf{c})\rrbracket\llbracket A(\textbf{a}_b,\textbf{a}_c,\mathsf{a}) = C(\textbf{c}_a,\textbf{c}_b,\mathsf{c})\rrbracket\notag\\
&=& \sum_{\textbf{a}_b}  \pp_S(\textbf{a}_b|\mathsf{abc})\sum_{\textbf{c}_a,\textbf{c}_b} \pp_S(\textbf{c}_a,\textbf{c}_b|\mathsf{abc})\sum_{\textbf{a}_c}\llbracket \textbf{a}_c = \textbf{A}_c^{S}(\textbf{a}_b,\textbf{c}_a,\textbf{c}_b,\mathsf{a},\mathsf{c})\rrbracket\llbracket A(\textbf{a}_b,\textbf{a}_c,\mathsf{a}) = C(\textbf{c}_a,\textbf{c}_b,\mathsf{c})\rrbracket\notag\\
&=& \sum_{\textbf{a}_b}\pp_S(\textbf{a}_b|\mathsf{abc}) \sum_{\textbf{c}_a,\textbf{c}_b} \pp_S(\textbf{c}_a,\textbf{c}_b|\mathsf{abc})\llbracket A(\textbf{a}_b,{\bf A}_c^{S}(\textbf{a}_b,\textbf{c}_a,\textbf{c}_b,\mathsf{a},\mathsf{c}),\mathsf{a}) = C(\textbf{c}_a,\textbf{c}_b,\mathsf{c})\rrbracket.\label{e:splits}
\end{eqnarray}
The value of the inner sum $\sum_{\textbf{c}_a,\textbf{c}_b}$ will vary depending on the particular $\textbf{a}_b$ fixed by the outer sum, and there will thus be an optimal value $\textbf{a}_b^*$ that maximizes the inner sum. This optimal value is not necessarily unique; the inner sum could even be the same value for all choices of $\textbf{a}_b$, but this is consistent with there being at least one optimal value. 

Now we are ready to define a new strategy $S'$, which only differs from $S$ for Alice, and only when her setting choice $X$ is $\mathsf{a}$. Effectively, Alice's modification is to ``pretend" to observe $\textbf{a}_b^*$ for $\textbf{A}_b$ and work through her $S$ decision in this manner, where $\textbf{a}_b^*$ is the chosen value maximizing the inner sum in \eqref{e:splits}.\footnote{$\textbf{a}_b^*$ is analogous to $r^*_v$ appearing in Eq. (3) of Ref.~\cite{chao:2017}. Note however that $r^*_v$ is a value of an introduced random variable intended to parametrize the randomness of bipartite nonsignaling nonlocal subsystems whereas $\textbf{a}_b^*$ is an output of the bipartite nonlocal signaling subsystems proper.} Formally, we re-define Alice's decision tree in $S'$ as follows: starting at the left of Figure \ref{f:strategy} and working to the right, all branches are the same as those of $S$ except for when a $A_b^i$ location is encountered. At such a location, the path according to $S$ splits into two paths based on the value of $A_b^i$; to obtain the tree for $S'$, re-define the remainder of \textit{both} paths moving rightward to be what $S$ does when the $i$th value of $\textbf{a}_b^*$ is observed for $A_b^i$. 
Repeat this substitution for all $A_b^i$ locations on the decision tree, moving from left to right. When this process is complete, define the $S'$ function for choosing the final output as $A(\textbf{a}_b^*,\textbf{a}_c,\mathsf{a})$ -- i.e., substitute the fixed string $\textbf{a}_b^*$ for whatever is Alice's actually observed value of $\textbf{A}_b$ into the function $A(\cdot,\cdot,\cdot)$ of strategy $S$. 

So under $S'$, we have that for given values of $\textbf{A}_b=\textbf{a}_b$, $\textbf{C}_a=\textbf{c}_a$, $\textbf{C}_b=\textbf{c}_b$, and $Z=z\in\{\mathsf{c},\mathsf{c}\textnormal{\textquotesingle}\}$, then if $X=\mathsf{a}$ the determined string $\textbf{A}_c$ in Figure \ref{f:ctoa} will populate according to the old function $\textbf{A}_c^{S}(\textbf{a}_b^*,\textbf{c}_a,\textbf{c}_b,\mathsf{a},z)$ with the fixed string $\textbf{a}_b^*$ as input instead of $\textbf{a}_b$. Hence the following relationship holds between $\textbf{A}_c^{S'}(\cdot)$ and $\textbf{A}_c^{S}(\cdot)$, which are we recall the functions that determine $\textbf{A}_c$ from $\textbf{A}_b,\textbf{A}_a,\textbf{A}_b$ under strategies $S'$ and $S$, respectively:
\begin{equation}\label{e:removeab}
\textbf{A}_c^{S'}(\textbf{A}_b,\textbf{C}_a,\textbf{C}_b,\mathsf{a},z)=\textbf{A}_c^{S}(\textbf{a}_b^*,\textbf{C}_a,\textbf{C}_b,\mathsf{a},z).
\end{equation}
Thus when Alice's setting $X$ is $\mathsf{a}$, the output of the function $\textbf{A}_c^{S'}(\cdot)$ is independent of $\textbf{A}_b$. This enables us to express Alice's new final output function $A'(\cdot)$ under strategy $S'$ as a function $A'({\bf A}_b,{\bf A}_c,X)$ that does not depend on ${\bf A}_b$ when $X=\mathsf{a}$, satisfying 
\begin{equation}\label{e:outremoveab}
A'({\bf A}_b,{\bf A}_c,\mathsf{a}) =A(\textbf{a}_b^*,\textbf{A}_c^{S}(\textbf{a}_b^*,\textbf{C}_a,\textbf{C}_b,\mathsf{a},Z),\mathsf{a}).
\end{equation}
A key implication of \eqref{e:outremoveab} is that when $X=\mathsf{a}$, Alice's final output $A$ is a function of (only) $\textbf{C}_a,\textbf{C}_b$, and $Z$. Returning to \eqref{e:splits}, we have, for every value of $\textbf{a}_b$,
\begin{multline*}
\sum_{\textbf{c}_a,\textbf{c}_b} \pp_S(\textbf{c}_a,\textbf{c}_b|\mathsf{abc})\llbracket A(\textbf{a}_b,{\bf A}_c^{S}(\textbf{a}_b,\textbf{c}_a,\textbf{c}_b,\mathsf{a},\mathsf{c}),\mathsf{a}) = C(\textbf{c}_a,\textbf{c}_b,\mathsf{c})\rrbracket\\
\le \sum_{\textbf{c}_a,\textbf{c}_b} \pp_{S}(\textbf{c}_a,\textbf{c}_b|\mathsf{abc})\llbracket A(\textbf{a}_b^*,\textbf{A}_c^{S}(\textbf{a}_b^*,\textbf{c}_a,\textbf{c}_b,\mathsf{a},\mathsf{c}),\mathsf{a}) = C(\textbf{c}_a,\textbf{c}_b,\mathsf{c})\rrbracket 
\end{multline*}
which follows from the manner in which $\textbf{a}_b^*$ was chosen. Now by \eqref{e:threeindep1}, $\pp_{S}(\textbf{c}_a,\textbf{c}_b|\mathsf{abc})=\pp_{S'}(\textbf{c}_a,\textbf{c}_b|\mathsf{abc})=2^{-2n}$ and $\pp_{S}(\textbf{a}_b|\mathsf{abc})=\pp_{S'}(\textbf{a}_b|\mathsf{abc})=2^{-n}$, so we can write that \eqref{e:splits} is less than or equal to
\begin{eqnarray*}
&&\sum_{\textbf{a}_b,\textbf{c}_a,\textbf{c}_b}  \pp_{S'}(\textbf{a}_b|\mathsf{abc})\pp_{S'}(\textbf{c}_a,\textbf{c}_b|\mathsf{abc})\llbracket A(\textbf{a}_b^*,\textbf{A}_c^{S}(\textbf{a}_b^*,\textbf{c}_a,\textbf{c}_b,\mathsf{a},\mathsf{c}),\mathsf{a}) = C(\textbf{c}_a,\textbf{c}_b,\mathsf{c})\rrbracket \\
&=& \sum_{\textbf{a}_b,\textbf{a}_c,\textbf{c}_a,\textbf{c}_b}  \pp_{S'}(\textbf{a}_b|\mathsf{abc})\pp_{S'}(\textbf{c}_a,\textbf{c}_b|\mathsf{abc})\llbracket \textbf{a}_c = \textbf{A}_c^{S}(\textbf{a}_b^*,\textbf{c}_a,\textbf{c}_b,\mathsf{a},\mathsf{c})\rrbracket\llbracket A'(\textbf{a}_b,\textbf{a}_c,\mathsf{a}) = C(\textbf{c}_a,\textbf{c}_b,\mathsf{c})\rrbracket\notag\\
&=& \sum_{\textbf{a}_b,\textbf{a}_c,\textbf{c}_a,\textbf{c}_b}  \pp_{S'}(\textbf{a}_b|\mathsf{abc})\pp_{S'}(\textbf{c}_a,\textbf{c}_b|\mathsf{abc})\llbracket \textbf{a}_c = \textbf{A}_c^{S'}(\textbf{a}_b,\textbf{c}_a,\textbf{c}_b,\mathsf{a},\mathsf{c})\rrbracket\llbracket A'(\textbf{a}_b,\textbf{a}_c,\mathsf{a}) = C(\textbf{c}_a,\textbf{c}_b,\mathsf{c})\rrbracket\notag\\
&=& \sum_{\textbf{a}_b,\textbf{a}_c,\textbf{c}_a,\textbf{c}_b} \pp_{S'}(\textbf{a}_b,\textbf{a}_c,\textbf{c}_a,\textbf{c}_b|\mathsf{abc})\llbracket A'(\textbf{a}_b,\textbf{a}_c,\mathsf{a}) = C(\textbf{c}_a,\textbf{c}_b,\mathsf{c})\rrbracket\notag\\
&=& \pp_{S'}(A=C|\mathsf{abc})\notag,
\end{eqnarray*}
where we used \eqref{e:removeab} in the third line and \eqref{e:parametrizing} in the fourth line. Hence
\begin{equation}\label{e:partone}
\pp_{S}(A=C|\mathsf{abc})\le \pp_{S'}(A=C|\mathsf{abc}).
\end{equation}
This also holds if the conditioner is $\mathsf{ab\textnormal{\textquotesingle}c}$, as the argument above did not depend on the choice of $Y$.

\subsection{Comparing the performance of $S'$ and $S$ when $Z=\mathsf{c}\textnormal{\textquotesingle}$}

The first part of the proof, up until now, showed that the newly-defined strategy $S'$ has an at-least-as-good ``win" probability as $S$ for the first and third bracketed terms in \eqref{e:uncompact}, corresponding to the setting $X=\mathsf{a}$ when $Z=\mathsf{c}$. Strategy $S'$ involves Alice changing her behavior on setting $\mathsf{a}$, so the second part of the proof is to analyze how $S'$ performs compared to $S$ for the other configurations involving setting $\mathsf{a}$: the second and fourth bracketed terms in \eqref{e:uncompact} when $Z=\mathsf{c}\textnormal{\textquotesingle}$. The goal is to show that there is a bound to how much ``worse'' $S'$ can do on these other settings. The argument is as follows, which we perform for setting $\mathsf{abc\textnormal{\textquotesingle}}$, but which also works for setting $\mathsf{ab\textnormal{\textquotesingle}c\textnormal{\textquotesingle}}$. The steps below are similar to the manipulations following Eq.~(4) of \cite{chao:2017}:
\begin{eqnarray*}
\pp_{S'}(A\ne B|\mathsf{abc\textnormal{\textquotesingle}})&=& \pp_{S'}(A\ne B|\mathsf{abc}) \quad\quad \quad \quad\quad \quad\quad\quad \quad\text{(by no-signaling; see \eqref{e:nosig})}\\ 
&\le& \pp_{S'}(\{A\ne C\} \cup \{C \ne B\} |\mathsf{abc}) \\
&\le& \pp_{S'}(A\ne C |\mathsf{abc})+\pp_{S'}(C \ne B|\mathsf{abc}) \\
&\le& \pp_{S}(A\ne C |\mathsf{abc})+\pp_{S'}(C \ne B|\mathsf{abc}),\quad\quad\quad\quad\quad\text{(by \eqref{e:partone})}
\end{eqnarray*}
where we used the event relationship $\{A\ne B\}\subseteq\{A\ne C\} \cup \{C \ne B\}$ and the union bound in the second and third lines. Continuing with the second term in the last line above, we have
\begin{eqnarray*}
\pp_{S'}(C \ne B|\mathsf{abc})&=&\pp_{S}(C \ne B|\mathsf{abc}) \quad\quad\quad\quad\quad\quad\quad\quad\quad\quad \text{($S=S'$ for $B$ and $C$)}\\
&\le& \pp_{S}(\{C \ne A\} \cup \{A \ne B\} |\mathsf{abc}) \\
&\le& \pp_{S}(C \ne A |\mathsf{abc})+\pp_{S}(A \ne B |\mathsf{abc})\\
&\le& \pp_{S}(C \ne A |\mathsf{abc})+\pp_{S}(A \ne B |\mathsf{abc\textnormal{\textquotesingle}}) \quad \quad \quad \text{(by no-signaling),}\\
\end{eqnarray*} 
where the first equation follows from the independence of Charlie and Bob's joint distribution from changes to Alice's strategy; see the discussion surrounding \eqref{e:nosig}. Thus we obtain the final inequality
\begin{equation}\label{e:parttwo}
\pp_{S'}(A\ne B|\mathsf{abc\textnormal{\textquotesingle}}) \le 2\pp_{S}(A\ne C |\mathsf{abc})+\pp_{S}(A \ne B |\mathsf{abc\textnormal{\textquotesingle}}).
\end{equation}
This complements \eqref{e:partone} in constraining the possibilities for an increase of $E(F)$ when moving from $S$ to $S'$, analogous to the expressions at the bottom of page 12 of Ref.~\cite{chao:2017}. Since the above inequality also holds if we substitute $\mathsf{b}\textnormal{\textquotesingle}$ for $\mathsf{b}$, we can apply \eqref{e:parttwo} to \eqref{e:uncompact} to write
\begin{eqnarray}\label{e:nkotb}
E_S(F) &=& \frac{1}{8}\big[2\pp_S(A\ne C|\mathsf{abc}) +\pp_S(A\ne B|\mathsf{abc}\textnormal{\textquotesingle})+2\pp_S(A\ne C|\mathsf{ab\textnormal{\textquotesingle}c})\notag \\
&&+\pp_S(A\ne B|\mathsf{ab\textnormal{\textquotesingle}c\textnormal{\textquotesingle}})\times1+\pp_S(A\oplus B\oplus C=0|\mathsf{a\textnormal{\textquotesingle}bc\textnormal{\textquotesingle}})+\pp_S(A\oplus B\oplus C=1|\mathsf{a\textnormal{\textquotesingle}b\textnormal{\textquotesingle}c\textnormal{\textquotesingle}})\big]\notag\\
&\ge& \frac{1}{8}\big[\pp_{S'}(A\ne B|\mathsf{abc\textnormal{\textquotesingle}}) +\pp_{S'}(A\ne B|\mathsf{ab\textnormal{\textquotesingle}c\textnormal{\textquotesingle}})\notag\\
&&+\pp_{S'}(A\oplus B\oplus C=0|\mathsf{a\textnormal{\textquotesingle}bc\textnormal{\textquotesingle}})+\pp_{S'}(A\oplus B\oplus C=1|\mathsf{a\textnormal{\textquotesingle}b\textnormal{\textquotesingle}c\textnormal{\textquotesingle}})\big].
\end{eqnarray}
Note that equality holds for the terms with $X=\mathsf{a}\textnormal{\textquotesingle}$ because strategies $S$ and $S'$ only differ on setting $X=\mathsf{a}$. An interpretation of \eqref{e:nkotb}  is that a shift of Alice's $X=\mathsf{a}$ strategy that improves the $A\ne C$ terms will worsen the $A\ne B$ terms, but only to a certain degree such that the new strategy's performance on $F$ for a restricted subset of setting configurations lower bounds the original strategy's expected $F$ value. As we will see in the next section, the particular manner in which $S'$ was obtained from $S$ implies independence of Alice and Bob such that $P(A=a,B=b|\mathsf{ay})=P(A=a|\mathsf{a})P(B=b|\mathsf{y})$ for $\mathsf{y}\in\{\mathsf{b},\mathsf{b}\textnormal{\textquotesingle}\}$; this independence condition allows the derivation of a non-trivial lower bound for \eqref{e:nkotb}.

\subsection{Modifying the strategy $S'$ to an even simpler strategy $S''$}

To bound \eqref{e:nkotb} from below, we show how to move from the strategy $S'$ to a strategy $S''$ for which Alice has a constant outcome for $A$ on setting $X=\mathsf{a}$, and where $P_{S'}(A=B|\mathsf{ay}\mathsf{c}\textnormal{\textquotesingle}) \le P_{S''}(A=B|\mathsf{ay}\mathsf{c}\textnormal{\textquotesingle})$ for both choices of $\mathsf{y}\in \{\mathsf{b},\mathsf{b}\textnormal{\textquotesingle}\}$. We will then see that it is impossible to simultaneously make all four probabilities in \eqref{e:nkotb} arbitrarily close to zero with a nonsignaling distribution where Alice has a constant outcome. To construct $S''$ recall that on strategy $S'$, Alice's output function $A'(\textbf{A}_b,\textbf{A}_c,X)$ doesn't depend on $\textbf{A}_b$ when $X=\mathsf{a}$. To emphasize this, we write Alice's output function as $A'(\textbf{A}_c,\mathsf{a})$ and use a version of \eqref{e:threeindep1} to see that
\begin{flalign*}
\pp_{S'}(A=a,B=b|\mathsf{abc}\textnormal{\textquotesingle})&= \sum_{\textbf{a}_c,\textbf{b}_a,\textbf{b}_c} \pp_{S'}(\textbf{a}_c,\textbf{b}_a,\textbf{b}_c|\mathsf{abc}\textnormal{\textquotesingle})\llbracket A'(\textbf{a}_c,\mathsf{a}) =a\rrbracket\llbracket B(\textbf{b}_a,\textbf{b}_c,\mathsf{c}\textnormal{\textquotesingle})=b\rrbracket\notag\\
&= \sum_{\textbf{a}_c,\textbf{b}_a,\textbf{b}_c}  \pp_{S'}(\textbf{a}_c|\mathsf{abc}\textnormal{\textquotesingle})\pp_{S'}(\textbf{b}_a,\textbf{b}_c|\mathsf{abc}\textnormal{\textquotesingle})\llbracket A'(\textbf{a}_c,\mathsf{a}) =a\rrbracket\llbracket B(\textbf{b}_a,\textbf{b}_c,\mathsf{c}\textnormal{\textquotesingle})=b\rrbracket\notag\\
&= \sum_{\textbf{a}_c} \pp_{S'}(\textbf{a}_c|\mathsf{abc}\textnormal{\textquotesingle})\llbracket A'(\textbf{a}_c,\mathsf{a}) =a\rrbracket\sum_{\textbf{b}_a,\textbf{b}_c}  \pp_{S'}(\textbf{b}_a,\textbf{b}_c|\mathsf{abc}\textnormal{\textquotesingle})\llbracket B(\textbf{b}_a,\textbf{b}_c,\mathsf{c}\textnormal{\textquotesingle})=b\rrbracket\notag\\
&= \pp_{S'}(A=a|\mathsf{abc}\textnormal{\textquotesingle})\pp_{S'}(B=b|\mathsf{abc}\textnormal{\textquotesingle}),
\end{flalign*}
and so Alice's and Bob's distributions are independent for this setting configuration. This independence similarly holds for the configuration $\mathsf{ab\textnormal{\textquotesingle}c\textnormal{\textquotesingle}}$, so for the first two bracketed terms in \eqref{e:nkotb} we can write
\begin{eqnarray*}
\sum_{\mathsf{y} \in \{\mathsf{b}, \mathsf{b}\textnormal{\textquotesingle}\}}\pp_{S'}(A\ne B|\mathsf{ayc\textnormal{\textquotesingle}})
&=&\sum_{\mathsf{y} \in \{\mathsf{b}, \mathsf{b}\textnormal{\textquotesingle}\}}\sum_{k\in \{\text{0},\text{+}\}}\pp_{S'}(A=k, B=\neg k|\mathsf{ayc\textnormal{\textquotesingle}})\\
&=&\sum_{k\in \{\text{0},\text{+}\}}\sum_{\mathsf{y} \in \{\mathsf{b}, \mathsf{b}\textnormal{\textquotesingle}\}}\pp_{S'}(A=k|\mathsf{ayc\textnormal{\textquotesingle}})\pp_{S'}(B=\neg k|\mathsf{ayc\textnormal{\textquotesingle}})\\
&=&\sum_{k\in \{\text{0},\text{+}\}}\pp_{S'}(A=k|\mathsf{ayc\textnormal{\textquotesingle}})\sum_{\mathsf{y} \in \{\mathsf{b}, \mathsf{b}\textnormal{\textquotesingle}\}}\pp_{S'}(B=\neg k|\mathsf{ayc\textnormal{\textquotesingle}}).
\end{eqnarray*}
From this, we see that Alice can always decrease this expression (or at least keep it the same) by moving to a strategy with a fixed output for $A$ on setting $\mathsf{a}$: specifically, always choose the value of $k\in \{\text{0},\text{+}\}$ that induces the minimum of the two possible values of $\sum_{\mathsf{y} \in \{\mathsf{b}, \mathsf{b}\textnormal{\textquotesingle}\}}\pp_{S'}(B=\neg k|\mathsf{ayc\textnormal{\textquotesingle}})$. (This choice is analogous to $c_z$ in the equation preceding Claim 8.3 of Ref.~\cite{chao:2017}.) If we define $S''$ to be the strategy that does this for Alice on setting $\mathsf{a}$, and is otherwise the same as $S'$ (and $S$), we  can continue from \eqref{e:nkotb} to get
\begin{eqnarray}\label{e:nkotb2}
E_S(F) &\ge& \frac{1}{8}\big[\pp_{S'}(A\ne B|\mathsf{abc\textnormal{\textquotesingle}}) +\pp_{S'}(A\ne B|\mathsf{ab\textnormal{\textquotesingle}c\textnormal{\textquotesingle}})\notag\\
&&+\pp_{S'}(A\oplus B\oplus C=0|\mathsf{a\textnormal{\textquotesingle}bc\textnormal{\textquotesingle}})+\pp_{S'}(A\oplus B\oplus C=1|\mathsf{a\textnormal{\textquotesingle}b\textnormal{\textquotesingle}c\textnormal{\textquotesingle}})\big]\notag\\
&\ge& \frac{1}{8}\big[\pp_{S''}(A\ne B|\mathsf{abc\textnormal{\textquotesingle}}) +\pp_{S''}(A\ne B|\mathsf{ab\textnormal{\textquotesingle}c\textnormal{\textquotesingle}})\notag\\
&&+\pp_{S''}(A\oplus B\oplus C=0|\mathsf{a\textnormal{\textquotesingle}bc\textnormal{\textquotesingle}})+\pp_{S''}(A\oplus B\oplus C=1|\mathsf{a\textnormal{\textquotesingle}b\textnormal{\textquotesingle}c\textnormal{\textquotesingle}})\big].
\end{eqnarray}

\subsection{A bound for $E_S(F)$, and quantum violation}

The last step is to note that a strategy like $S''$ with a fixed output on a given setting is limited in its ability to win certain nonlocal games, and use this fact to bound the overall win probability of the original strategy $S$ using \eqref{e:nkotb2}. Indeed, the bracketed quantity 
\begin{equation}\label{e:bracketedquantity}
\pp_{S''}(A\ne B|\mathsf{abc\textnormal{\textquotesingle}}) +\pp_{S''}(A\ne B|\mathsf{ab\textnormal{\textquotesingle}c\textnormal{\textquotesingle}})+\pp_{S''}(A\oplus B\oplus C=0|\mathsf{a\textnormal{\textquotesingle}bc\textnormal{\textquotesingle}})+\pp_{S''}(A\oplus B\oplus C=1|\mathsf{a\textnormal{\textquotesingle}b\textnormal{\textquotesingle}c\textnormal{\textquotesingle}})
\end{equation}
is effectively a variant of the CHSH Bell quantity for Alice and Bob, with Charlie's setting $Z$ fixed as $\mathsf{c\textnormal{\textquotesingle}}$. It can be determined with standard arguments that if Alice has a fixed output on setting $X=\mathsf{a}$, and the probability distribution is nonsignaling, then the sum of these conditional probabilities must be at least one; see Appendix \ref{s:lb} for a proof. This leads to 
\begin{equation}\label{e:finalresult}
E_S(F)\ge \frac{1}{8}.\hspace{1cm}\Box
\end{equation}
\medskip
 
The bound can be violated by a quantum mechanics. Using the state and measurements defined in Proposition 5.1 of Ref.~\cite{chao:2017}, which here are
\begin{equation}
\ket{\psi} = \frac{\ket{000}+\ket{111}}{\sqrt 2}, \quad \mathsf{a}=\mathsf{c}=\sigma^z,\quad \mathsf{a}\textnormal{\textquotesingle}=\mathsf{c}\textnormal{\textquotesingle}=\sigma^x,\quad\mathsf{b}= \frac{\sigma^z+\sigma^x}{\sqrt 2}, \quad \mathsf{b}\textnormal{\textquotesingle}= \frac{\sigma^z-\sigma^x}{\sqrt 2},
\end{equation}
we obtain the following settings-conditional probabilities, using the shorthand 
$\mathcal C = (1/4)\cos^2\left(\pi/8 \right)$ and $\mathcal S = (1/4)\sin^2\left(\pi/8\right)$: 
\begin{center}
\begin{tabular}{|c|cccccccc|}
\hline
& \multicolumn{8}{c|}{Outcomes $ABC$}\\
  & +++&++0&+0+&+00&0++&0+0&00+&000\\
\hline
$\mathsf{abc}$&$2\mathcal C$&$0$&$2 \mathcal S$&$0$&$0$&$2\mathcal S$&$0$&$2\mathcal C$\\
$\mathsf{abc\textnormal{\textquotesingle}}$&$\mathcal C$&$\mathcal C$&$\mathcal S$&$\mathcal S$&$\mathcal S$&$\mathcal S$&$\mathcal C$&$\mathcal C$\\
$\mathsf{ab\textnormal{\textquotesingle}c}$&$2\mathcal C$&$0$&$2 \mathcal S$&$0$&$0$&$2\mathcal S$&$0$&$2\mathcal C$\\
$\mathsf{ab\textnormal{\textquotesingle}c\textnormal{\textquotesingle}}$&$\mathcal C$&$\mathcal C$&$\mathcal S$&$\mathcal S$&$\mathcal S$&$\mathcal S$&$\mathcal C$&$\mathcal C$\\
$\mathsf{a\textnormal{\textquotesingle}bc}$&$\mathcal C$&$\mathcal S$&$\mathcal S$&$\mathcal C$&$\mathcal C$&$\mathcal S$&$\mathcal S$&$\mathcal C$\\
$\mathsf{a\textnormal{\textquotesingle}bc\textnormal{\textquotesingle}}$&$\mathcal C$&$\mathcal S$&$\mathcal S$&$\mathcal C$&$\mathcal S$&$\mathcal C$&$\mathcal C$&$\mathcal S$\\
$\mathsf{a\textnormal{\textquotesingle}b\textnormal{\textquotesingle}c}$&$\mathcal C$&$\mathcal S$&$\mathcal S$&$\mathcal C$&$\mathcal C$&$\mathcal S$&$\mathcal S$&$\mathcal C$\\
$\mathsf{a\textnormal{\textquotesingle}b\textnormal{\textquotesingle}c\textnormal{\textquotesingle}}$&$\mathcal S$&$\mathcal C$&$\mathcal C$&$\mathcal S$&$\mathcal C$&$\mathcal S$&$\mathcal S$&$\mathcal C$\\
\hline
\end{tabular}
\end{center}
Recalling that the setting probabilities are each 1/8, we can compute the expectation for the above quantum strategy as follows:
\begin{equation*}
E_Q(B)=2\mathcal S \simeq 0.07322 < \frac{1}{8}
\end{equation*}

\section{Conclusion}\label{s:conclusion}

We have demonstrated the existence of a joint distribution for the outputs of a network of PR boxes shared among three parties, whose uniqueness is guaranteed by a few well motivated principles. We have shown that the distribution obeys various no-signaling properties and certain functional dependencies between different sets of PR box outputs. We used these attributes to rigorously prove that a version of the inequality described in Ref.~\cite{chao:2017} must be obeyed by all tripartite behaviors that can be induced by underlying networks of PR boxes; in turn, the results of Ref.~\cite{forster:2011} imply that the inequality is obeyed more generally by tripartite behaviors induced by underlying networks of arbitrary bipartite nonlocal nonsignaling systems with access to shared local randomness. The version of the inequality presented here applies readily to an experimental setup enforcing space-like separation of the measuring parties, and the inequality can be violated robustly by an appropriate quantum state and measurement. 


An experimental violation of the inequality \eqref{e:finalresult} can be interpreted as a demonstration of a version of genuine tripartite nonlocality that differs from previous definitions of the concept \cite{svetlichny:1987,bancal:2013,gallego:2012,dutta:2020}. Interestingly, a perspective on genuine tripartite nonlocality similar to the one discussed in the Introduction can be found in Definition 1 of a recent work of Schmid \textit{et al.}~\cite{schmid:2020}. There, the shared bipartite subsystems are restricted to be quantum states, which is different from the scenario studied in this paper: bipartite nonsignaling nonlocal behaviors need not be quantum-achievable in general -- indeed the PR box is not -- but quantum states also can be subject to entangled measurements in a manner that PR boxes cannot \cite{short:2006,barrett:2007}, leading to different types of joint probability distributions. Exploring the possible modification of constraints like \eqref{e:finalresult} under different restrictions on the shared bipartite resources is a potential avenue of future research. Indeed, recent works in network nonlocality \cite{renou:2019,gisin:2020} have derived constraints for a similar ``triangle network'' of bipartite-only nonlocal resources; the scenario of these works differs from the present one however in disallowing a source of local randomness shared among all three parties.

Future work should also provide a fuller characterization of the set of tripartite behaviors admitting an underlying PR-box-network model. These models cannot properly accommodate conflicting two-party correlation conditions like $A= B$ versus $A=C$ along with three-party correlations like $A\otimes B \otimes C = k$ in differing measurement configurations. To discover new noise-robust constraints, the approach of the arguments of Section 3 can be explored in other scenarios with similar characteristics, such as the scenario of Scarani \cite{scarani:9.2.2}. A more general question is whether the set of PR-box-network-simulable correlations is a polytope, like the classical/local realist set and the no-signaling set. If it is a polytope, the determination of its facet inequalities would help analyze whether the inequality \eqref{e:finalresult} can be improved upon. The techniques of this paper can provide a foundation for proving results along these lines; specifically, if the conditions beyond no-signaling that are obeyed by PR-box networks can be precisely formulated, arguments of the form of Section 2 can be used for rigorously proving them.

Unfortunately, it is not clear whether the framework presented here can be extended in a straightforward manner to $n$-party scenarios for $n>3$ (such as, for instance, deriving a constraint in the four-party setting obeyed by all distributions that can be induced by networks of tripartite-only nonsignaling nonlocal subsystems). This is because there is no ``$n$-partite PR box'' known to simulate all other $n$-partite nonsignaling nonlocal correlations for values of $n$ greater than 2. Notwithstanding, the three-party scenario of this paper will be the first multipartite scenario accessible to experimental setups of increasing sophistication. Determining the tightest inequalities and strongest quantum violations for this setting will streamline the path to a potential future experiment demonstrating a phenomenon transcending bipartite nonsignaling nonlocal behavior.

\medskip

\noindent\textbf{Acknowledgements.} The author thanks Stefano Pironio for pointing out the relevance of Ref.~\cite{chao:2017}, as well as anonymous referees for helpful feedback. This work was partially supported by Louisiana Board of Regents award LEQSF (2019-22)-RD-A-27 and NSF award 1839223.


\appendix

\section{Problematic Joint Distributions of Boxes}\label{s:sigpr}

Consider a scenario where Alice and Bob share two binary input, binary output boxes, in which Alice always sees the same output from both boxes -- i.e., with probability 1/2 she sees a ``1" from box 1 and 2, and with probability 1/2 she sees a ``0" from box 1 and 2.  Alice's marginal distribution of each box conforms with \eqref{e:PRprobs}. Of course, this does not comport with what we would expect from two PR boxes which should not be correlated with each other this way. Indeed, such a joint distribution would generate signaling effects if Alice puts inputs into her boxes sequentially: if Alice wanted to send the digit 0 (1) to Bob, she could observe her first box and feed that output (the opposite of that output) into the second box as its input. Then provided Bob inputs 1 to the second box, he will observe Alice's message bit. Indeed, if Alice's marginal distribution for the second box is anything other than uniform, conditioned on the first box's output, she will be able to send information to Bob through the above strategy, if imperfectly. The avoidance of such signaling pathologies clarifies the role of \eqref{e:stipulation} and \eqref{e:stipulation2}. 

\section{Lower Bounding the Quantity in \eqref{e:bracketedquantity}}\label{s:lb}

Here we show that for a nonsignaling distribution $\pp$, the quantity in \eqref{e:bracketedquantity} is lower bounded by $1$ if Alice has the fixed outcome ``+" on setting $\mathsf{a}$. The argument can be directly modified to prove the scenario where Alice has the fixed outcome 0 by performing the exchange $\text{0} \leftrightarrow \text{+}$ for each instance of Alice's and Charlie's outcome in the expressions below.

If Alice has the fixed outcome + on setting $\mathsf{a}$, then for the setting configuration $\mathsf{ab}\textnormal{\textquotesingle}\mathsf{c}\textnormal{\textquotesingle}$ the sum of all the probabilities where $A$ is equal to ``+" must equal one:
\begin{equation}\label{e:pirategame}
\pp(\text{+++}|\mathsf{ab}\textnormal{\textquotesingle}\mathsf{c}\textnormal{\textquotesingle})+\pp(\text{++0}|\mathsf{ab}\textnormal{\textquotesingle}\mathsf{c}\textnormal{\textquotesingle})+\pp(\text{+0+}|\mathsf{ab}\textnormal{\textquotesingle}\mathsf{c}\textnormal{\textquotesingle})^L+\pp(\text{+00}|\mathsf{ab}\textnormal{\textquotesingle}\mathsf{c}\textnormal{\textquotesingle})^L=1
\end{equation}
Here we have introduced a shorthand where, for instance, $\pp(\text{+0+}|\mathsf{ab}\textnormal{\textquotesingle}\mathsf{c}\textnormal{\textquotesingle})$ equals $\pp(A=\text{+},B=\text{0},C=\text{+}|\mathsf{ab}\textnormal{\textquotesingle}\mathsf{c}\textnormal{\textquotesingle})$, and the last two terms are marked with an $L$ superscript to highlight that they contribute to the sum in \eqref{e:bracketedquantity}. We show that the first two terms are upper bounded by a sum of probabilities that all contribute to \eqref{e:bracketedquantity} in a non-overlapping manner. First, we have
\begin{eqnarray}
\pp(\text{+++}|\mathsf{ab}\textnormal{\textquotesingle}\mathsf{c}\textnormal{\textquotesingle})
&=&\pp(\text{+++}|\mathsf{ab}\textnormal{\textquotesingle}\mathsf{c}\textnormal{\textquotesingle})+\pp(\text{0++}|\mathsf{ab}\textnormal{\textquotesingle}\mathsf{c}\textnormal{\textquotesingle})\notag\\
&=&\pp(\text{+++}|\mathsf{a}\textnormal{\textquotesingle}\mathsf{b}\textnormal{\textquotesingle}\mathsf{c}\textnormal{\textquotesingle})^L+\pp(\text{0++}|\mathsf{a}\textnormal{\textquotesingle}\mathsf{b}\textnormal{\textquotesingle}\mathsf{c}\textnormal{\textquotesingle})\notag\\
&\le&\pp(\text{+++}|\mathsf{a}\textnormal{\textquotesingle}\mathsf{b}\textnormal{\textquotesingle}\mathsf{c}\textnormal{\textquotesingle})^L+\pp(\text{0++}|\mathsf{a}\textnormal{\textquotesingle}\mathsf{b}\textnormal{\textquotesingle}\mathsf{c}\textnormal{\textquotesingle})+\pp(\text{00+}|\mathsf{a}\textnormal{\textquotesingle}\mathsf{b}\textnormal{\textquotesingle}\mathsf{c}\textnormal{\textquotesingle})\notag\\
&=&\pp(\text{+++}|\mathsf{a}\textnormal{\textquotesingle}\mathsf{b}\textnormal{\textquotesingle}\mathsf{c}\textnormal{\textquotesingle})^L+\pp(\text{0++}|\mathsf{a}\textnormal{\textquotesingle}\mathsf{b}\mathsf{c}\textnormal{\textquotesingle})^L+\pp(\text{00+}|\mathsf{a}\textnormal{\textquotesingle}\mathsf{b}\mathsf{c}\textnormal{\textquotesingle})\notag\\
&\le&\pp(\text{+++}|\mathsf{a}\textnormal{\textquotesingle}\mathsf{b}\textnormal{\textquotesingle}\mathsf{c}\textnormal{\textquotesingle})^L+\pp(\text{0++}|\mathsf{a}\textnormal{\textquotesingle}\mathsf{b}\mathsf{c}\textnormal{\textquotesingle})^L+\pp(\text{00+}|\mathsf{a}\textnormal{\textquotesingle}\mathsf{b}\mathsf{c}\textnormal{\textquotesingle})+\pp(\text{+0+}|\mathsf{a}\textnormal{\textquotesingle}\mathsf{b}\mathsf{c}\textnormal{\textquotesingle})\notag\\
&=&\pp(\text{+++}|\mathsf{a}\textnormal{\textquotesingle}\mathsf{b}\textnormal{\textquotesingle}\mathsf{c}\textnormal{\textquotesingle})^L+\pp(\text{0++}|\mathsf{a}\textnormal{\textquotesingle}\mathsf{b}\mathsf{c}\textnormal{\textquotesingle})^L+\pp(\text{00+}|\mathsf{a}\mathsf{b}\mathsf{c}\textnormal{\textquotesingle})+\pp(\text{+0+}|\mathsf{a}\mathsf{b}\mathsf{c}\textnormal{\textquotesingle})^L\notag\\
&=&\pp(\text{+++}|\mathsf{a}\textnormal{\textquotesingle}\mathsf{b}\textnormal{\textquotesingle}\mathsf{c}\textnormal{\textquotesingle})^L+\pp(\text{0++}|\mathsf{a}\textnormal{\textquotesingle}\mathsf{b}\mathsf{c}\textnormal{\textquotesingle})^L+\pp(\text{+0+}|\mathsf{a}\mathsf{b}\mathsf{c}\textnormal{\textquotesingle})^L\label{e:apb1}
\end{eqnarray}
where the first equality holds because the inserted term is zero, the inequalities hold because probabilities are nonnegative, the last equality holds because the removed term is zero, and the other equalities hold due to no-signaling. We similarly upper bound the second term of \eqref{e:pirategame} as follows:
\begin{eqnarray}
\pp(\text{++0}|\mathsf{ab}\textnormal{\textquotesingle}\mathsf{c}\textnormal{\textquotesingle})
&=&\pp(\text{++0}|\mathsf{ab}\textnormal{\textquotesingle}\mathsf{c}\textnormal{\textquotesingle})+\pp(\text{0+0}|\mathsf{ab}\textnormal{\textquotesingle}\mathsf{c}\textnormal{\textquotesingle})\notag\\
&=&\pp(\text{++0}|\mathsf{a}\textnormal{\textquotesingle}\mathsf{b}\textnormal{\textquotesingle}\mathsf{c}\textnormal{\textquotesingle})+\pp(\text{0+0}|\mathsf{a}\textnormal{\textquotesingle}\mathsf{b}\textnormal{\textquotesingle}\mathsf{c}\textnormal{\textquotesingle})^L\notag\\
&\le&\pp(\text{++0}|\mathsf{a}\textnormal{\textquotesingle}\mathsf{b}\textnormal{\textquotesingle}\mathsf{c}\textnormal{\textquotesingle})+\pp(\text{+00}|\mathsf{a}\textnormal{\textquotesingle}\mathsf{b}\textnormal{\textquotesingle}\mathsf{c}\textnormal{\textquotesingle})+\pp(\text{0+0}|\mathsf{a}\textnormal{\textquotesingle}\mathsf{b}\textnormal{\textquotesingle}\mathsf{c}\textnormal{\textquotesingle})^L\notag\\
&=&\pp(\text{++0}|\mathsf{a}\textnormal{\textquotesingle}\mathsf{b}\mathsf{c}\textnormal{\textquotesingle})^L+\pp(\text{+00}|\mathsf{a}\textnormal{\textquotesingle}\mathsf{b}\mathsf{c}\textnormal{\textquotesingle})+\pp(\text{0+0}|\mathsf{a}\textnormal{\textquotesingle}\mathsf{b}\textnormal{\textquotesingle}\mathsf{c}\textnormal{\textquotesingle})^L\notag\\
&\le&\pp(\text{++0}|\mathsf{a}\textnormal{\textquotesingle}\mathsf{b}\mathsf{c}\textnormal{\textquotesingle})^L+\pp(\text{+00}|\mathsf{a}\textnormal{\textquotesingle}\mathsf{b}\mathsf{c}\textnormal{\textquotesingle})+\pp(\text{000}|\mathsf{a}\textnormal{\textquotesingle}\mathsf{b}\mathsf{c}\textnormal{\textquotesingle})+\pp(\text{0+0}|\mathsf{a}\textnormal{\textquotesingle}\mathsf{b}\textnormal{\textquotesingle}\mathsf{c}\textnormal{\textquotesingle})^L\notag\\
&=&\pp(\text{++0}|\mathsf{a}\textnormal{\textquotesingle}\mathsf{b}\mathsf{c}\textnormal{\textquotesingle})^L+\pp(\text{+00}|\mathsf{a}\mathsf{b}\mathsf{c}\textnormal{\textquotesingle})^L+\pp(\text{000}|\mathsf{a}\mathsf{b}\mathsf{c}\textnormal{\textquotesingle})+\pp(\text{0+0}|\mathsf{a}\textnormal{\textquotesingle}\mathsf{b}\textnormal{\textquotesingle}\mathsf{c}\textnormal{\textquotesingle})^L\notag\\
&=&\pp(\text{++0}|\mathsf{a}\textnormal{\textquotesingle}\mathsf{b}\mathsf{c}\textnormal{\textquotesingle})^L+\pp(\text{+00}|\mathsf{a}\mathsf{b}\mathsf{c}\textnormal{\textquotesingle})^L+\pp(\text{0+0}|\mathsf{a}\textnormal{\textquotesingle}\mathsf{b}\textnormal{\textquotesingle}\mathsf{c}\textnormal{\textquotesingle})^L.\label{e:apb2}
\end{eqnarray}
Combining \eqref{e:pirategame}, \eqref{e:apb1}, and \eqref{e:apb2}, we have
\begin{eqnarray*}
1 &\le& \pp(\text{+++}|\mathsf{a}\textnormal{\textquotesingle}\mathsf{b}\textnormal{\textquotesingle}\mathsf{c}\textnormal{\textquotesingle})+\pp(\text{0++}|\mathsf{a}\textnormal{\textquotesingle}\mathsf{b}\mathsf{c}\textnormal{\textquotesingle})+\pp(\text{+0+}|\mathsf{a}\mathsf{b}\mathsf{c}\textnormal{\textquotesingle})+\pp(\text{++0}|\mathsf{a}\textnormal{\textquotesingle}\mathsf{b}\mathsf{c}\textnormal{\textquotesingle})\\
&&+\pp(\text{+00}|\mathsf{a}\mathsf{b}\mathsf{c}\textnormal{\textquotesingle})+\pp(\text{0+0}|\mathsf{a}\textnormal{\textquotesingle}\mathsf{b}\textnormal{\textquotesingle}\mathsf{c}\textnormal{\textquotesingle})+\pp(\text{+0+}|\mathsf{ab}\textnormal{\textquotesingle}\mathsf{c}\textnormal{\textquotesingle})+\pp(\text{+00}|\mathsf{ab}\textnormal{\textquotesingle}\mathsf{c}\textnormal{\textquotesingle})\\
&\le&\pp(A\ne B|\mathsf{abc\textnormal{\textquotesingle}}) +\pp(A\ne B|\mathsf{ab\textnormal{\textquotesingle}c\textnormal{\textquotesingle}})+\pp(A\oplus B\oplus C=0|\mathsf{a\textnormal{\textquotesingle}bc\textnormal{\textquotesingle}})+\pp(A\oplus B\oplus C=1|\mathsf{a\textnormal{\textquotesingle}b\textnormal{\textquotesingle}c\textnormal{\textquotesingle}}).\quad\Box
\end{eqnarray*}

\bibliographystyle{unsrt}
\bibliography{metabib}

\end{document}